\documentclass{aa}  

\usepackage{graphicx}
\usepackage{subfigure}
\usepackage{txfonts}
\usepackage{amsmath}
\usepackage{appendix}
\usepackage[hyperfootnotes=false]{hyperref}

\begin{document}

   \title{Resonance Capture and Stability Analysis for Planet Pairs under Type I Disk Migration}

   \author{Linghong Lin\inst{1} \and Beibei Liu\inst{1 \star} \and Zekai Zheng\inst{1,2} 
          }

   \institute{Institute for Astronomy, School of Physics, Zhejiang University, Hangzhou 310027,  China\\
              \email{[llh\_astro; bbliu]@zju.edu.cn}
              \and
               Department of Physics, National University of Singapore, Singapore 117542, Singapore\\
              \email{zekai77@u.nus.edu}
             }

   \date{}

 
  \abstract
   {We present a theoretical framework for investigating a two-planet system undergoing convergent type I migration in a protoplanetary disk. Our study identifies the conditions for resonant capture and subsequent dynamical stability. 
   By deriving analytical criteria for general $j$:$j-1$ first-order mean-motion resonances
(MMRs) applicable to planet pairs with arbitrary mass ratios, we validate these predictions through N-body simulations.
 The key results are demonstrated in $\tau_{\rm m}$-$\tau_{\rm m}/\tau_{e}$ plots, where $\tau_{\rm m}$ and $\tau_{e}$ are the timescales of the angular momentum and eccentricity damping, respectively.
 Specifically, we determine which combinations of orbital damping timescales allow for capture into resonance, showing that too fast migration or too strong eccentricity damping inhibit successful capture.
 After capture, the subsequent evolution can be classified into three regimes: stable trap, overstable trap and escape. Importantly, resonant capture always remains stable when the inner planet significantly outweighs the outer one. In contrast, when the mass of the inner planet is lower than or comparable to that of the outer planet, the system transitions from the stable to overstable trap, and eventually escapes the resonance,  as the relative strength of eccentricity damping to migration ($\tau_{\rm m}/\tau_{e}$) decreases.
   }

   \keywords{planets and satellites: dynamical evolution and stability -- celestial mechanics -- methods: analytical}

   \maketitle
%

\section{Introduction}
Low-mass planets embedded in protoplanetary disks interact with surrounding disk gas and undergo type I orbital migration \citep{Goldreich&Tremaine1980,Ward+1997, Tanaka+2002,Kley&Nelson2012, Baruteau2013, Paardeekoper2023}. Resonance capture can be a natural outcome of a planet pair that undergoes convergent migration \citep{Lee&Peale2002,Papaloizou&Szuszkiewicz2005}. The key question is to understand 
 under which conditions this capture can occur and its following dynamical evolution.
 
 The classical pendulum model intuitively shows that the trapping holds when the time planet migrates across the resonance width is longer than the libration period in the resonance \citep{1999ssd..book.....M, 2013ApJ...775...34O,Liu&Zhang2015}, which is akin to the adiabatic condition described by \cite{2015MNRAS.451.2589B}. The eccentricity damping from the gas disk is neglected in these studies.  Later, \cite{Kajtazi2023} demonstrated through simulations that eccentricity damping qualitatively alters the capture conditions. Incorporating eccentricity damping, \cite{2023MNRAS.522..828H} derived a new criterion by analyzing the resonant equilibrium state under disk-driven migration within a restricted three-body framework. Meanwhile, \cite{2023ApJ...946L..11B} derived an equivalent criterion using Hamiltonian theory and extended the analysis to the two-planet pairs with arbitrary mass ratios.
 
  On the other hand, \cite{2014AJ....147...32G} first investigated the long-term stability of the resonant planet pair after capture in a gaseous disk. Their study revealed that under certain conditions, the resonant equilibrium state can become overstable, where the planets may eventually escape from resonance. Building on this, \cite{2015ApJ...810..119D} conducted a similar analysis using Hamiltonian theory and extended the work to the unrestricted three-body system. \cite{2017MNRAS.468.3223X} further expanded the framework to second-order mean motion resonances. These studies highlighted that the mass ratio of the planet pair is a crucial parameter in determining whether the system is overstable. \cite{2022ApJ...925...38N} applied the overstable libration to account for the high eccentricity and large amplitude of the resonant angle observed in TOI-216b. Recently, \cite{2024A&A...686A.277A} verified this overstable resonant trap using hydrodynamical simulations. They found that the outcome is related to disk properties, such as the gas mass and turbulent viscosity. 

  In this paper, we develop a unified framework for a two-planet pair 
 that undergoes convergent type I migration and gets trapped at the first-order mean motion resonance ($j$:$j-1$ MMR). We investigate the conditions under which planets can be captured in resonances and examine the subsequent stability properties. The key analytical criteria are derived for planets with arbitrary mass ratios, and the outcomes are summarized in a $\tau_{\rm m}{-} \tau_{\rm m}/\tau_{e}$ parameter space. 
  
  The paper is organized as follows. In Section \ref{sec:analysis},  we present an analytical model that describes the dynamics of planets in resonant configurations. In Section \ref{sec:simulation} we perform N-body simulations and compare the results with analytical predictions. Section \ref{sec:discuss} outlines potential caveats, and Section \ref{sec:conclusion} summarizes our key results.
\section{Dynamics of resonance capture and stability}
\label{sec:analysis}
This section discusses the dynamical evolution of a two-planet pair in a gaseous disk through the convergent type-I migration.  
We consider two planets of masses $m_{\rm i}$ and $m_{\rm o}$ orbiting a central star of $M_{\star}$ on coplanar orbits.  Their normalized masses and the mass ratio between the inner and outer ones are $\mu_{\rm i,o} {= }m_{\rm i,o} / M_{\rm \star}$ and $q{=}m_{\rm i}/m_{\rm o}$,  respectively. For notation, the subscripts ‘i’ and ‘o’ refer to the corresponding quantities of the inner and outer planets. 

We present the analytical criteria for resonance trapping and stability in Section \ref{sec:criteria} and Section \ref{sec:stability}, respectively. A unified picture is summarized in Section \ref{sec:picture}.  Resonance crossing in the context of divergent migration is also discussed in Section \ref{sec:div_mig}. 

\subsection{Resonance trapping criteria}
\label{sec:criteria} 
The dissipative effect of the planet-disk interaction can be parameterized by \citep{2014MNRAS.443..568T, Ataiee&Kely2021, Pichierri2023}
\begin{subequations} \label{disk_ef}
\begin{eqnarray}
      \frac{1}{L}\frac{dL}{dt} & = & -\frac{1}{\tau_{\rm m}}, \\
   \frac{1}{a}\frac{da}{dt} & = & -\frac{1}{\tau_{\rm a}} =   -\frac{2}{\tau_{\rm m}} + \frac{2e^{2}}{1-e^{2}}\frac{1}{\tau_{e}},
    \label{eq:da_dt2} \\
      \frac{1}{e}\frac{de}{dt} & = & -\frac{1}{\tau_{e}} \label{de_dt},
\end{eqnarray} 
\end{subequations}
where the orbital elements $L$, $a$, $n$, $e$, $\lambda$, $\varpi$, denote the angular momentum, semimajor axis, mean motion, eccentricity, mean longitude and longitude of pericenter of the planet, $\tau_{\rm m}$, $\tau_{\rm a}$ and $\tau_{\rm e}$ represent the damping timescales for orbital angular momentum, semi-major axis and eccentricity, respectively. It is worth noting that in the limit of small eccentricity, the angular momentum and orbital decay timescales can be linked through $\tau_{\rm m}{\approx} 2\tau_{\rm a}$.

\begin{figure}
     \centering
     \includegraphics[width=0.5\textwidth]{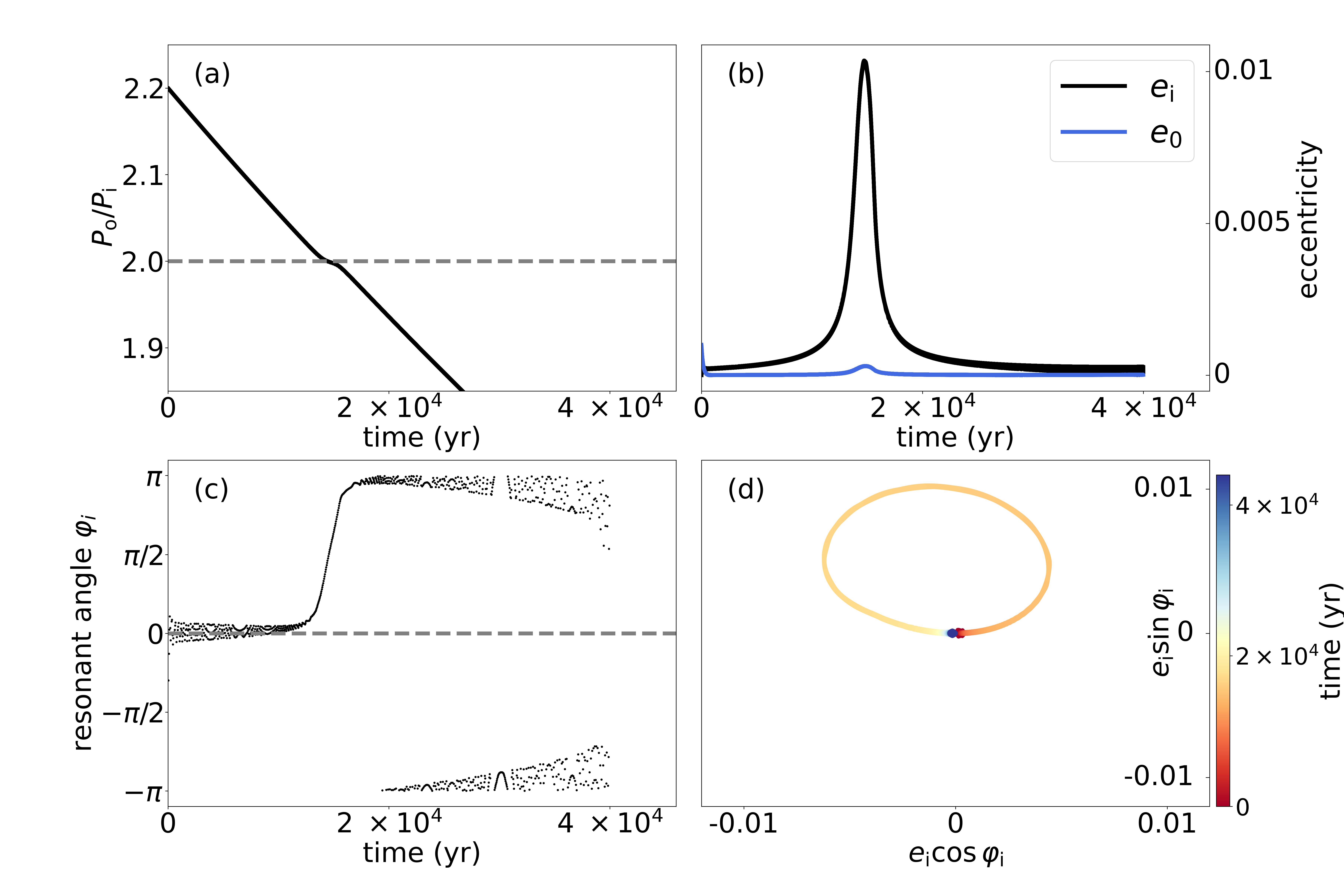}
     \caption{
     Resonance crossing of a two-planet system with convergent migration. The plot shows the time evolution of (a) the outer-to-inner planet period ratio, (b) the planet eccentricities, (c) the resonant angle, and (d) the phase curve. The planet and disk parameters are $m_{\rm i}{=}1\ M_{\oplus}$, $m_{\rm o}{=}10\ M_{\oplus}$, $\tau_{\rm m} {= }2.2 \times 10^5 $ yr and $\tau_{\rm m} / \tau_{e} {=} 3 \times 10^3$.
     The planets directly cross the $2$:$1$ resonance at $t{\simeq}6\times 10^{3}$ yr, while the inner planet's eccentricity briefly excites to $0.01$ before rapid decay. 
     During the crossing, the resonant angle of the inner planet $\varphi_{\rm i}$ jumps from $0$ to $\pi$ and subsequently oscillates with growing amplitude\protect\footnotemark.
     In phase space, the planet’s trajectory is not captured by the resonant equilibrium. Instead, it circulates and eventually evolves toward the origin, corresponding to a low-eccentricity, non-resonant state. See \protect\url{https://github.com/llh-astro/resonance/raw/main/gif/no_trap.gif}
}
     \label{fig:res_not_trap}
\end{figure}

\footnotetext[1]{After crossing, the angle may appear to librate due to the short-period terms of disturbing function, but this does not indicate true resonance \citep{Delisle2012}. The same applies to Fig.~\ref{fig:esc}.}
For a $j$:$j-1$ MMR, the resonant angles and the time-averaged disturbing functions for the inner and outer planets are given by \citep{1999ssd..book.....M} $\varphi_{\rm i} {=} j\lambda_{\rm o} 
                  + (1-j)\lambda_{\rm i}
                  -\varpi_{\rm i}$,
$\varphi_{\rm o} {=} j\lambda_{\rm o} 
                  + (1-j)\lambda_{\rm i}
                  -\varpi_{\rm o}$,              
and  $  R_{\rm i} { =}  Gm_{\rm o}/a_{\rm o} (R_{\rm D}^{\rm sec} + R_{\rm D}^{\rm res} + \alpha R_{\rm E})$,  $R_{\rm o}  {=}  Gm_{\rm i}/a_{\rm o} (R_{\rm D}^{\rm sec} + R_{\rm D}^{\rm res} + R_{\rm I}/\alpha^{2})$, 
where $\alpha {=} a_{\rm i}/a_{\rm o}$ is the two planets' semi-major axis ratio, $R_{\rm D}^{\rm sec}$ and $R_{\rm D}^{\rm res}$ are the secular and resonant terms, $R_{\rm E}$ and $R_{\rm I}$ are the indirect part due to an external and internal perturber.

We assume small eccentricities of the planets, and expand the disturbing functions to the first order in $e_{\rm i}$ and $e_{\rm o}$, which gives $R_{\rm D}^{\rm sec} { =} 0$, $R_{\rm D}^{\rm res} {= } e_{\rm i}f_{\rm d,i}\cos{\varphi_{\rm i}} + e_{\rm o}f_{\rm d,o}\cos{\varphi_{\rm o}}$,$R_{\rm E} {= } -2\delta_{\rm j,2}e_{\rm o}\cos{\varphi_{\rm o}}$, and $R_{\rm I} { =} -\frac{1}{2}\delta_{\rm j,2}e_{\rm o}\cos{\varphi_{\rm o}}$,
where $\delta_{\rm j,2}$ is the usual Kronecker symbol. Detailed expressions of $f_{\rm d,i}$ and $f_{\rm d,o}$ can be found in Table 8.1 of \cite{1999ssd..book.....M}.

Taking orbital decay and eccentricity damping into account, Lagrange's equations can be written as
\begin{subequations}
\begin{align}
     \dot{n_{\rm i}} &     =  -3(j-1) \mu_{\rm o} n_{\rm i}^{2} \alpha 
                          ( e_{\rm i}f_{\rm d,i}\sin{\varphi_{\rm i}} + e_{\rm o}f_{\rm d,o}^{'}\sin{\varphi_{\rm o}} )
                          +\frac{3n_{\rm i}}{\tau_{\rm m \rm,i}}  + \frac{3n_{\rm i}e_{\rm i}^{2}}{\tau_{e,\rm i}} \label{dot_ni},\\
   \dot{n_{\rm o}}    &   =  3j \mu_{\rm i} n_{\rm o}^{2} 
                          ( e_{\rm i}f_{\rm d,i}\sin{\varphi_{\rm i}} + e_{\rm o}f_{\rm d,o}^{'}\sin{\varphi_{\rm o}} )
                          +\frac{3n_{\rm o}}{\tau_{\rm m \rm,o}} + \frac{3n_{\rm o}e_{\rm o}^{2}}{\tau_{e,\rm o}} \label{dot_no},\\
  \dot{e_{\rm i}}      &  =  -\mu_{\rm o} n_{\rm i} \alpha f_{\rm d,i} \sin{\varphi_{\rm i}}
                          - \frac{e_{\rm i}}{\tau_{e,\rm i}}   \label{dot_ei}, \\ 
   \dot{e_{\rm o}}     &  =  -\mu_{\rm i} n_{\rm o} f_{\rm d,o}^{'} \sin{\varphi_{\rm o}}
                          - \frac{e_{\rm o}}{\tau_{e,\rm o}}   \label{dot_eo}, \\
   \dot{\varpi_{\rm i}} & =  \mu_{\rm o} n_{\rm i} \alpha f_{\rm d,i} \frac{\cos{\varphi_{\rm i}}}{e_{\rm i}}
   \label{dot_wi},\\
   \dot{\varpi_{\rm o}}  & =  \mu_{\rm i} n_{\rm o} f_{\rm d,o}^{'}     \frac{\cos{\varphi_{\rm o}}}{e_{\rm o}}
   \label{dot_wo},
\end{align}
\end{subequations}
where $f_{\rm d,o}^{'} {=} f_{\rm d,o} - 2\alpha \delta_{j,2}$. For instance, $f_{\rm d,i}{=}{-}1.19$, $f_{\rm d,o}^{'}{=}0.43$ at the $2$:$1$ MMR.

The orbital parameters of the planets reach the equilibrium state at resonances such that
$\dot{n_{\rm i}}/n_{\rm i}  {=} \dot{n_{\rm o}}/n_{\rm o}$, 
$\dot{e_{\rm i}}  {= } \dot{e_{\rm o}} {=} 0$, 
$\dot{\varpi_{\rm i}} {=}\dot{\varpi_{\rm o}} $. By substituting $\dot{n_{\rm i}}/n_{\rm i}  {=} \dot{n_{\rm o}}/n_{\rm o}$ into Lagrange's equation, Eqs. \eqref{dot_ni} and \eqref{dot_no} can be combined as 
\begin{equation}
\begin{aligned}
   &  -[ (j-1) \mu_{\rm o} n_{\rm i} \alpha
    + j \mu_{\rm i} n_{\rm o} ]
    ( e_{\rm i} f_{\rm d,i} \sin{\varphi_{\rm i}}
    + e_{\rm o} f_{\rm d,o}^{'} \sin{\varphi_{\rm o}} )\\
    &=  \frac{1}{\tau_{\rm m}}  + \frac{e_{\rm o}^2 - e_{\rm i}^{2}}{\tau_{e}},
    \label{n_eq}
    \end{aligned}
    \end{equation}
where we define that $\tau_{\rm m}^{-1} {=} \tau_{\rm m,o}^{-1} -\tau_{\rm m,i}^{-1}$ is the relative angular momentum damping rate for convergent migration, and $\tau_e=\tau_{e, \rm i}=\tau_{e, \rm o}$ represents the eccentricity damping timescale. In reality, the eccentricity damping rates of the two planets may vary, depending on disk properties and planet masses. For simplicity, we assume equal $\tau_e$ for both planets in our theoretical derivations and numerical simulations in the main text. We note, however, that our theoretical framework can be naturally extended to the case of unequal eccentricity damping. The expressions of our criteria for different eccentricity damping timescales are elaborated in Appendix C.

Substituting $\dot{e_{\rm i}}{=}\dot{e_{\rm o}}{=}0$ into Eqs. \eqref{dot_ei} and \eqref{dot_eo} to eliminate $\sin{\varphi_{\rm i,o}}$ and  $\dot{\varpi_{\rm i}} {=}\dot{\varpi_{\rm o}} $ into Eqs. \eqref{dot_wi} and \eqref{dot_wo} to eliminate either $e_{\rm i}$ or $e_{\rm o}$. Here we assume $\cos{\varphi_{\rm i}}=\cos{\varphi_{\rm o}}=1$. Under these conditions, the equilibrium eccentricities can be derived from Eq. \eqref{n_eq}, yielding  
\begin{subequations}
\begin{eqnarray}
   e_{\rm i,eq}^{2}                      & = &\frac{\tau_{e}}{\tau_{\rm m}}
                                           \frac{1}{1+q\sqrt{\alpha}}
                                           \frac{f_{\rm d,i}^{2}}{j f_{\rm d,i}^{2}+(j-1)f_{\rm d,o}^{'2}q\sqrt{\alpha}} \label{e_ieq}, \\
   e_{\rm o,eq}^{2}                      & = & \frac{\tau_{e}}{\tau_{\rm m}}
                                           \frac{1}{1+q\sqrt{\alpha}}
                                           \frac{f_{\rm d,o}^{'2} q^{2}\alpha}{j f_{\rm d,i}^{2}+(j-1)f_{\rm d,o}^{'2}q\sqrt{\alpha}} \label{e_oeq},
\end{eqnarray}
\label{e_eq}
\end{subequations}
where $e_{\rm o,eq}/e_{\rm i,eq}{=}{ q\sqrt{\alpha} |f_{\rm d,o}^{'}/f_{\rm d,i}|}$  (see \citealt{Lee2004, 2015ApJ...810..119D, Ramos2017, 2019MNRAS.482..530T}).

In the limit of the mass of the inner planet much lower than the outer one ($q {\ll} 1$), the eccentricity of the inner planet in Eq. \eqref{e_ieq}  can be simplified as 
\begin{equation}
   e_{\rm i,eq}  =  \sqrt{\frac{1}{j}
                      \frac{\tau_{e}}{\tau_{\rm m}}} \label{e_i_eq}.       
\end{equation}
 We note that due to differing signs of $f_{\rm d,i}$ and $f_{\rm d,o}^{'}$, $\dot{e_{\rm i}}{=}\dot{e_{\rm o}}{=}0$ requires that $\sin{\varphi_{\rm i}}{>}0$ and $\sin{\varphi_{\rm o}}{<}0$ for $j$:$j-1$ MMRs. Considering $e_{\rm o,eq}/e_{\rm i,eq}{=}{ q\sqrt{\alpha} |f_{\rm d,o}^{'}/f_{\rm d,i}|}$, we can derive $\sin{\varphi_{\rm i}} {=} -\sin{\varphi_{\rm o}}$, which implies $\varphi_{\rm i}-\varphi_{\rm o}{=}\pm \pi$. This means that the conjunctions between planet pairs in a mean motion resonance always occur at either pericentre or apocentre.

In particular, the right-hand side of Eq. \eqref{n_eq} represents the planet-disk interaction that leads to convergent orbital migration between two planets, while the left-hand side of Eq. \eqref{n_eq} represents their resonant gravitational interaction. When the effect of planet-disk interaction is strong enough to overcome the resonant interaction,  the planet pair cannot stay in resonance.  
The strong planet-disk interaction can be classified into two regimes: fast migration and strong eccentricity damping. In the first regime,  $\dot{a}/a$ in Eq. \eqref{n_eq} is driven by the angular momentum damping term ($\tau_{\rm m}^{-1} {\gg} e^2 \tau_{ e}^{-1}$). In the second regime, however, the contribution of the e-damping term becomes comparable to that of the angular momentum damping term. We will derive the resonance trapping conditions from these two regimes in the following subsections. 

\begin{figure}
      \centering
      \includegraphics[width=0.5\textwidth]{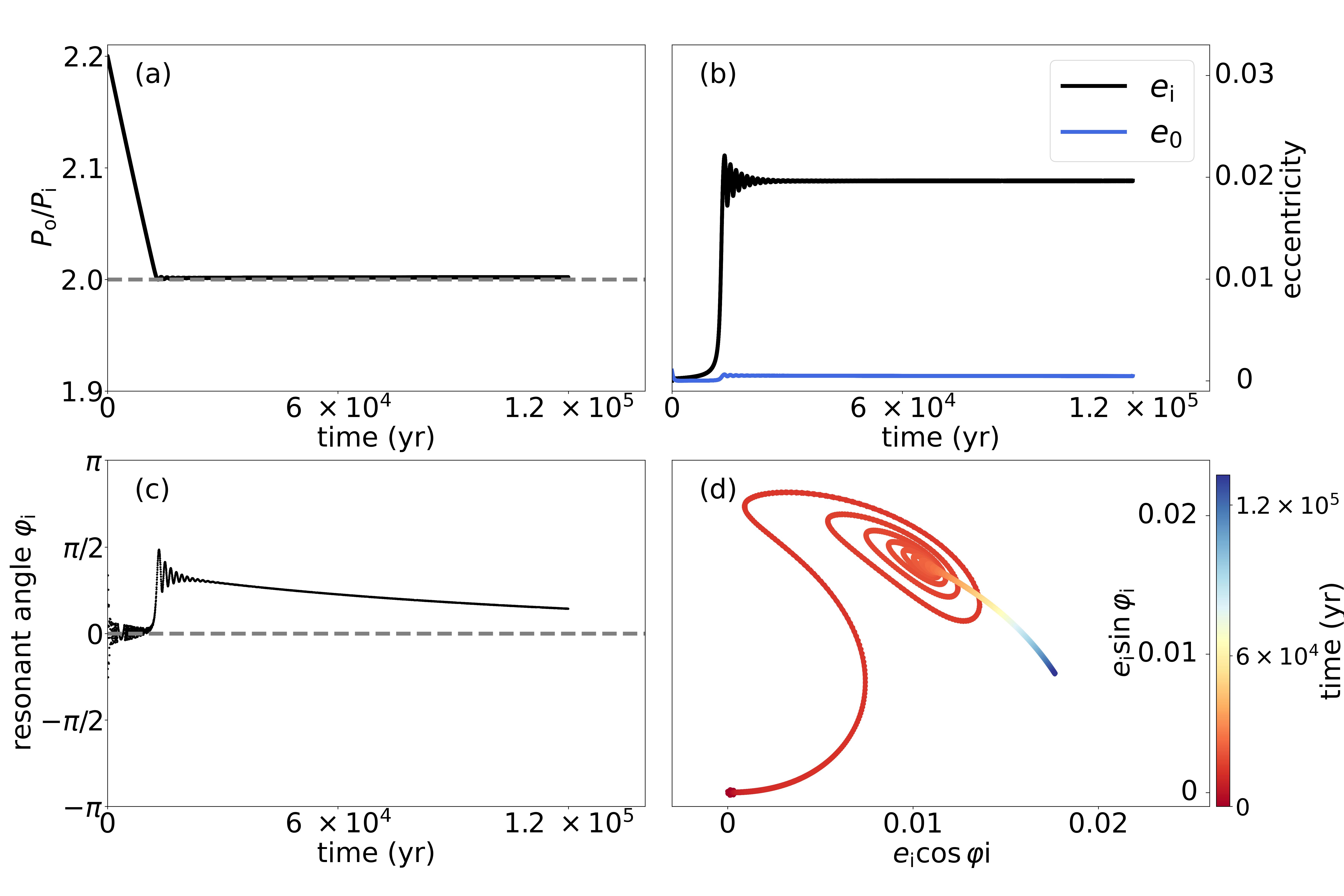}
      \caption{Stable resonant capture of a two-planet system with convergent migration. The planet and disk parameters are  $m_{\rm i}{=}1\ M_{\oplus}$, $m_{\rm o}{=}10\ M_{\oplus}$, $\tau_{\rm m} {= }2\ \times 10^5 $ yr and $\tau_{\rm m} / \tau_{e} {= }1200$.
      The planets get trapped in a $2$:$1$ MMR at $t{\simeq}10^{4}$ yr exciting the inner planet’s eccentricity to $0.02$. The resonant angle $\varphi_{\rm i}$ shifts from $0$ to $2\pi/3$ as the system transitions into the resonant state. It gradually decreases as convergent migration strengthens the resonant interaction, causing the libration center to drift slowly. In phase space, the trajectory converges toward the resonant equilibrium and co-evolves with it. We refer to this resonant capture at a stationary equilibrium as a stable trap. See \url{https://github.com/llh-astro/resonance/raw/main/gif/stable_trap.gif}
      }
      \label{fig:stab_trap}
\end{figure}
\subsubsection{Slow migration}
\label{sec:slow_mig}
Relatively slow migration is a necessary resonant capture condition. When the planets undergo fast convergent migration, they directly bypass the resonance before the eccentricities reach their equilibrium values. In this case $\dot{a}/a$ is driven by $\tau_{\rm m}$ but not $\tau_{e}$, and Eq. \eqref{n_eq} can be rewritten as 
\begin{equation}
    -[ (j-1) \mu_{\rm o} n_{\rm i} \alpha
    + j \mu_{\rm i} n_{\rm o} ]
    ( e_{\rm i} f_{\rm d,i}
    - e_{\rm o} f_{\rm d,o}^{'} ) =
    \frac{1}{\tau_{\rm m}},
    \label{slow_crit}
\end{equation}
 where the e-damping term is ignored. We adopt $\sin{\varphi_{\rm i}} {=} -\sin{\varphi_{\rm o}} {= }1$, so the left-hand side of Eq. \ref{slow_crit} represents the maximum strength of the resonant interaction. From the classical pendulum model (see Appendix A for the derivation), the maximum value of $e_{\rm i}f_{\rm d,i} - e_{\rm o}f_{\rm d,o}^{'} $  can be obtained as 
\begin{equation}
    e_{\rm i}f_{\rm d,i} - e_{\rm o}f_{\rm d,o}^{'} = 
    -\frac{3^{1/3}(\mu_{\rm o}n_{\rm i}\alpha f_{\rm d,i}^{2} + \mu_{\rm i}n_{\rm o}f_{\rm d,o}^{'2})^{2/3}}
    {[(j-1)^{2}\mu_{\rm o}n_{\rm i}^{2}\alpha + j^{2}\mu_{\rm i}n_{\rm o}^{2}]^{1/3}}. \label{pendul_ecce}
\end{equation}
If the migration rate is too rapid, the migration term in Eq. \ref{slow_crit} always exceeds the resonant term, and therefore the equilibrium state cannot be reached. Combing Eqs. \eqref{slow_crit} and \eqref{pendul_ecce},
the resonance trapping criterion of slow migration can be written as
\begin{equation}
    \tau_{\rm m}  > \frac{1}{[ (j-1) \mu_{\rm o} n_{\rm i} \alpha
    + j \mu_{\rm i} n_{\rm o} ]} 
    \frac{[(j-1)^{2}\mu_{\rm o}n_{\rm i}^{2}\alpha + j^{2}\mu_{\rm i}n_{\rm o}^{2}]^{1/3}}{3^{1/3}(\mu_{\rm o}n_{\rm i}\alpha f_{\rm d,i}^{2} + \mu_{\rm i}n_{\rm o}f_{\rm d,o}^{'2})^{2/3}}. \label{slow_mig}
\end{equation}
We further simplify the above criterion as
\begin{equation}
 \begin{cases}
 {\displaystyle   \tau_{\rm m} > 
                             \frac{1}{n_{\rm i}(3(j-1))^{1/3}} 
                             \frac{1}{(\mu_{\rm o} \alpha f_{\rm d,i})^{4/3}}  
                             \ \sim \mu_{\rm o}^{-4/3}
                             } \ \ \ \hfill  \mbox{at  $q \ll 1$}, \\
 {\displaystyle    \tau_{\rm m} > \frac{1}{n_{\rm o}(3j)^{1/3}} 
                             \frac{1}{(\mu_{\rm i} f_{\rm d,o}^{'})^{4/3}}  
                             \ \sim \mu_{\rm i}^{-4/3}  
                             } \ \ \  \hfill  \mbox{at $q \gg 1$}. \\
    
\end{cases}
\label{slow_crit: simp}
\end{equation}
This indicates that the threshold migration rate increases with the planet mass. Intuitively, the planet's resonant interaction becomes stronger as their masses increase. Therefore, the planets require faster migration in order to break the resonance.

\subsubsection{Weak eccentricity damping}

 When the planets migrate relatively slowly towards each other, they have time to reach the equilibrium state. Substituting the equilibrium eccentricity into the right side of Eq. \eqref{n_eq}, it turns out that the magnitude of the e-damping term is comparable to that of the angular momentum damping term. In this circumstance, the e-damping term becomes a key factor for resonance trapping. 

By substituting $e_{\rm i,eq}$ and $e_{\rm o,eq}$ into Eqs.\eqref{dot_ei} and \eqref{dot_eo}, $\sin{\varphi_{\rm i,o}}$ can be calculated. Naturally, the absolute values of $\sin{\varphi_{\rm i,o}}$ cannot exceed $1$. Therefore, by considering the limit of $\sin{\varphi_{\rm i}}{=}1$ and $\sin{\varphi_{\rm o}}{=}-1$, Eq.\eqref{n_eq} can be rewritten as  
\begin{equation}
       -[ (j-1) \mu_{\rm o} n_{\rm i} \alpha
    + j \mu_{\rm i} n_{\rm o} ]
    ( e_{\rm i} f_{\rm d,i}
    - e_{\rm o} f_{\rm d,o}^{'} ) =   \frac{1}{\tau_{\rm m}} + \frac{e_{\rm o}^2-e_{\rm i}^2}{\tau_{e}}.
    \label{e-damping}
\end{equation}  
If eccentricity damping is very strong, the orbital circularization leads to large $\dot{a}/a$, and the resonance can no longer hold. One example of resonance crossing is shown in Figure \ref{fig:res_not_trap}.  Hence, resonance trapping requires weak e-damping. Substituting Eq. \eqref{e-damping} with equilibrium values of $e_{\rm i}$ and $e_{\rm o}$ from Eqs. \eqref{e_ieq} and \eqref{e_oeq}, we obtain the resonance trapping criterion in the weak e-damping regime as
\begin{equation}
    \tau_{e}\tau_{\rm m} >  \frac{1}{(1+q\sqrt{\alpha})}
                       \frac{1}{[j f_{\rm d,i}^{2} + (j-1)f_{\rm d,o}^{',2} q\sqrt{\alpha}]}
                       \frac{1}{(\mu_{\rm o}n_{\rm i}\alpha)^{2}}. \label{weak damping}
\end{equation}
The expression agrees with the one obtained by \cite{2023ApJ...946L..11B}, who used a different yet equivalent analytical method. In the limit of $q {\ll} 1$, the trapping criterion can be simplified as 
\begin{equation}
    \tau_{e}\tau_{\rm m} > 
                       \frac{1}{j(\mu_{\rm o}n_{\rm i}\alpha f_{\rm d,i})^{2}},
\end{equation}
consistent with Eq. (25) of \cite{2023MNRAS.522..828H}.

Altogether, resonance trapping requires both relatively weak eccentricity damping and slow migration. Taking Eqs. \eqref{weak damping} and \eqref{slow_mig} together, we formulate the threshold condition for $\tau_{\rm m}$ and $\tau_{e}$ as 
  \begin{equation}
 \begin{cases}
 {\displaystyle   
 \tau_{e}\tau_{\rm m} > \frac{1}{(1+q\sqrt{\alpha})}
                       \frac{1}{[j f_{\rm d,i}^{2} + (j-1) f_{\rm d,o}^{',2} q\sqrt{\alpha}]}
                       \frac{1}{(\mu_{\rm o}n_{\rm i}\alpha)^{2}}   } \\
   {\displaystyle    }  
     \hfill  [\mbox{weak e-damping}],  \vspace{0.1cm}\\
 {\displaystyle       \tau_{\rm m} > \frac{1}{[ (j-1) \mu_{\rm o} n_{\rm i} \alpha
    + j \mu_{\rm i} n_{\rm o} ]} 
    \frac{[(j-1)^{2}\mu_{\rm o}n_{\rm i}^{2}\alpha + j^{2}\mu_{\rm i}n_{\rm o}^{2}]^{1/3}}{3^{1/3}(\mu_{\rm o}n_{\rm i}\alpha f_{\rm d,i}^{2} + \mu_{\rm i}n_{\rm o}f_{\rm d,o}^{'2})^{2/3}}  }\\
     {\displaystyle   } 
       \hfill  [\mbox{slow migration}].
\end{cases}
\label{eq:trap_criteria}
\end{equation}

    \begin{figure}
       \centering
       \includegraphics[width=0.5\textwidth]{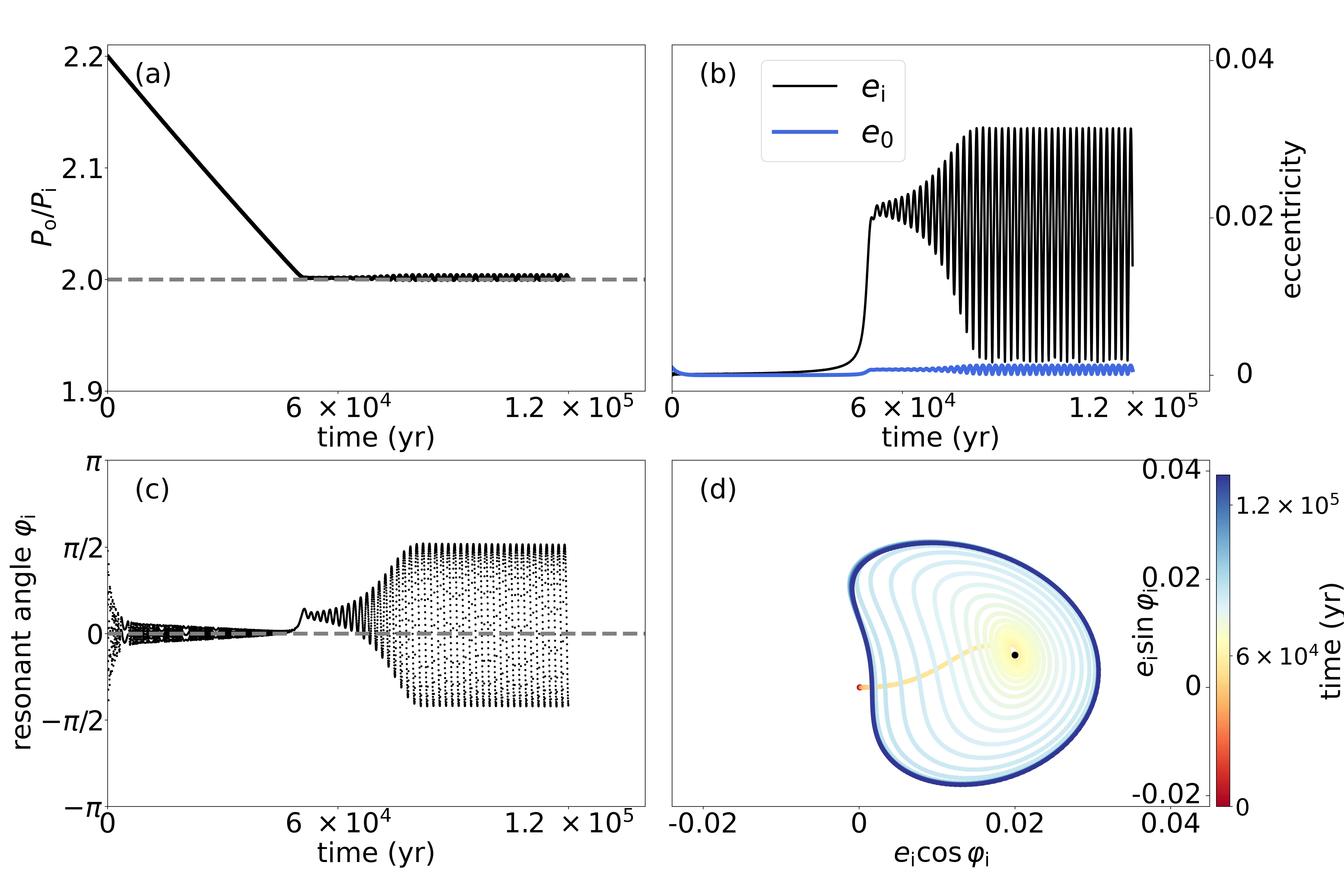}
       \caption{Overstable resonance capture of a two-planet system with convergent migration. 
       The planet and disk parameters are  $m_{\rm i}{=}1\ M_{\oplus}$, $m_{\rm o}{=}10\ M_{\oplus}$,  $\tau_{\rm m} {= }8 \times 10^5 $ yr and $\tau_{\rm m} / \tau_{e} {= }10^{3}$.
By  $t{\simeq}3\times 10^{4}$ yr, the planets become trapped in a $2$:$1$ MMR. The inner planet's eccentricity rapidly excites to $0.02$, followed by oscillations with significant amplitude, while the resonant angle exhibits similar behavior.
In phase space, the trajectory initially evolves toward a resonant equilibrium point (black dot) before stabilizing into a limit cycle around this point.     
We term this phenomenon$-$where the system remains trapped by librating around an equilibrium with the finite amplitude$-$an overstable trap. See \url{https://github.com/llh-astro/resonance/raw/main/gif/overstable_trap.gif}
       }
       \label{fig:overstab}
    \end{figure} 

\subsection{Post-evolution after resonant trapping}
\label{sec:stability}
    Here we assume that the planet pairs have satisfied the resonance trapping conditions in Section \ref{sec:criteria} and study their following dynamical evolution. The goal is to explore under which configuration the resonance state is temporary or long-term stable.   
    
    Our approach follows the analysis outlined in Section 2 of \cite{2017MNRAS.468.3223X} (also see  Chapter 8.8 of \citealt{1999ssd..book.....M}). We first focus on the situation that the mass of the inner planet is much lower than the outer one ($m_{\rm i}{\ll}m_{\rm o}$, $q{\ll}1$). In the absence of planet-disk interactions, the system's Hamiltonian $H$ near the $j$:$j-1$ MMR can be expressed as 
    \begin{equation}
      -H=\eta \Phi + \Phi^{2} -\Phi^{1/2}\cos{\varphi}, \label{H}
    \end{equation}
    where
      \begin{subequations}
    \begin{eqnarray}
      \varphi & = & j\lambda_{\rm o} 
                    + (1-j)\lambda_{\rm i} 
                    - \varpi_{\rm i},\\
      \Phi    & = & \left[\frac{3(j-1)^2}{f_{\rm d,i}\mu_{\rm o}\alpha_{\rm 0}} \right]^{2/3} 
                    \frac{1-\sqrt{1-e_{\rm i}^2}}{2} 
                    \sim \mu_{\rm o}^{2/3}e_{\rm i}^{2},\\
      \eta    & = & \left[\frac{1}{\sqrt{3}(j-1)f_{\rm d,i}\mu_{\rm o}} \right]^{2/3}
                  [(j-1)\alpha_{\rm 0}^{-2/3} - j\alpha_{\rm 0}^{5/6}].
    \end{eqnarray}
     \end{subequations}
     The first two parameters are a pair of conjugate coordinates and momentum, where $\varphi$ is the resonant angle and $\Phi$ represents the angular momentum deficit (AMD). The third parameter, $\eta$, is dimensionless and quantifies the deviation level from the exact resonance.  The orbital period ratio of the two planets is the exact integer ratio at $\eta{=}0$. The initial semi-major axis ratio of the two planets $ \alpha_{\rm 0} \equiv \alpha [1+(j-1)e_{\rm i}^2]$ is the conjugate momentum of a fast angle, which can be derived from Eq.8-92 of \cite{1999ssd..book.....M}. We note that in the absence of planet-disk interactions, the parameters $\eta$ and $\alpha_{0}$ remain constant.

\subsubsection{stability/overstability criterion}
   When accounting for the planet-disk interaction, the equations of motion of the system can be obtained from 
    \begin{equation}
    \dot{\Phi} {=} -\frac{\partial H}{\partial \varphi} + \frac{\partial \Phi}{\partial e}\frac{de}{dt} + \frac{\partial \Phi}{\partial a}\frac{da}{dt}, \  \  \  \dot{\varphi} {=} \frac{\partial H}{\partial \Phi}.
    \label{motion}
    \end{equation} 
   Since  $\Phi$ is independent of $a$ ($\alpha_0$ is a constant), the term involving 
   $\partial \Phi/\partial a=0$  can be neglected. By incorporating the angular momentum damping and eccentricity damping terms in Eq.\eqref{disk_ef}, Eq. \eqref{motion} can be derived as
   \begin{subequations}
    \begin{eqnarray}
        \dot{\Phi} &=& \Phi^{1/2}\sin{\varphi}-\frac{2}{\tau_e}\Phi, \\
        \dot{\varphi} &=& -\eta-2\Phi+\frac{1}{2\sqrt{\Phi}}\cos{\varphi}, \\
        \dot{\eta} &=& \frac{\partial \eta}{\partial \alpha_{0}} \frac{d \alpha_{0}}{d t} {=} -\frac{2\beta}{\tau_{\rm m}} + \frac{4j\Phi}{(j-1)\tau_e},
      \end{eqnarray}
    \end{subequations}
    where
    \begin{equation}
      \beta \equiv |\frac{\partial\eta}{\partial\ln\alpha_{\rm 0}}| = [\frac{1}{\sqrt{3}(j-1)f_{\rm d,i}\mu_{\rm o}}]^{2/3}\frac{3}{2}(j-1)^{5/9}j^{4/9},
    \end{equation}
    and $\Phi {=} \beta[(j-1)e_{\rm i}^2/2]$.
   
 The system reaches an equilibrium state at resonances when $\dot{\Phi}{=}0$, $\dot{\varphi}{=}0$, $\dot{\eta}{=}0$. Thus, the resonant parameters, to the lowest order, can be written as
      \begin{subequations}
    \begin{eqnarray}
      \Phi_{\rm eq}    &    =   &\frac{(j-1)\beta\tau_{e}}{2j\tau_{\rm m}},\\
      \varphi_{\rm eq} & \simeq & \frac{2}{\tau_{e}}\Phi_{\rm eq}^{1/2},\\
      \eta_{\rm eq}    & \simeq & \frac{1}{2\sqrt{\Phi_{\rm eq}}} 
                              - 2\Phi_{\rm eq}. \label{eq_eta}
    \end{eqnarray}
      \end{subequations}
    
    We then use the linear perturbation method to analyze the system's dynamics ($\Phi {=} \Phi_{\rm eq} +\delta\Phi$ etc.), and the corresponding linearized equations read 
\begin{equation}
    \resizebox{!}{0.75cm}{$
\begin{bmatrix}
\delta\dot{\Phi} \\
\delta\dot{\varphi} \\
\delta\dot{\eta}
\end{bmatrix}
=
\begin{bmatrix}
-\tau_{e}^{-1} & \Phi_{\rm eq}^{1/2} & 0\\
- (0.25\Phi^{-3/2}+2) & - \tau_{e}^{-1} & -1\\
4j \tau_{e}^{-1}(j-1)^{-1} & 0 & 0
\end{bmatrix}
\begin{bmatrix}
\delta{\Phi} \\
\delta{\varphi} \\
\delta{\eta}
\end{bmatrix}
$}.
\end{equation}   

We assume that the solutions have the following forms $\delta\Phi,\delta\varphi,\delta\eta \propto e^{\lambda t}$ and $\lambda$ is a complex number.  The characteristic equation can be written as
    \begin{equation}
      \lambda^3 +\frac{2}{\tau_{e}}\lambda^2 + \left(2\Phi_{\rm eq}^{1/2}+\frac{1}{4\Phi_{\rm eq}}\right)\lambda+\frac{2\beta}{\tau_{\rm m}\Phi_{\rm eq}^{1/2}}=0. \label{chara_eq} 
    \end{equation}
    
   We provide a general approach to solve the real part solutions of Eq. \eqref{chara_eq} as follows. Suppose that a cubic equation has a form of
 $  A\lambda^3 +B\lambda^2 + C\lambda+ D{=}0$
    where $A,B,C,D{>}0$. Assuming that $\lambda {=} \lambda_{\rm r} \pm i\lambda_{\rm i}$,  it can be proved that when $ D {<} BC/A$, the real part of eigenvalues $\lambda_r {<} 0$. This consequently means that the trapping is stable where the trajectory of Hamiltonian $H$ converges to the equilibrium (see an illustration in Figure \ref{fig:stab_trap}). Substituting $A,B,C,D$ with Eq. \eqref{chara_eq}, we obtain the stability criterion as
    \begin{equation}
      \frac{\tau_{\rm m}}{\tau_{e}} > \left(\frac{3}{\mu_{\rm o }}\right)^{2/3}\frac{1}{j}\left[\frac{(j-1)}{f_{\rm d,i}\alpha}\right]^{2/3} 
      \sim \mu_{\rm o}^{-2/3}. \label{stab_crit}
    \end{equation}
   Our result is consistent with Eq. (26) of \cite{2014AJ....147...32G}, where they constructed different forms of conserved quantities for derivations (see their Eq. 9).
    
     Our analysis can also be extended to the case when the mass of the outer planet is much lower than that of the inner one ($m_{\rm i} {\gg} m_{\rm o}$, $q{\gg}1$). In such a circumstance, we find that the real part of eigenvalues of the characteristic equation $\lambda_{\rm r} {<} 0$ always holds. Importantly, this means that once the planet pair gets trapped in the resonance, the trapping is always stable as long as $m_{\rm i} {\gg} m_{\rm o}$. This holds for the planet pairs with an equal eccentricity damping timescale. Detailed analysis for the planet pairs with different $\tau_{\rm e}$ can be found in \cite{2015ApJ...810..119D}. 

    \begin{figure} 
    \centering
    \subfigure{
        \includegraphics[width=0.4\textwidth]{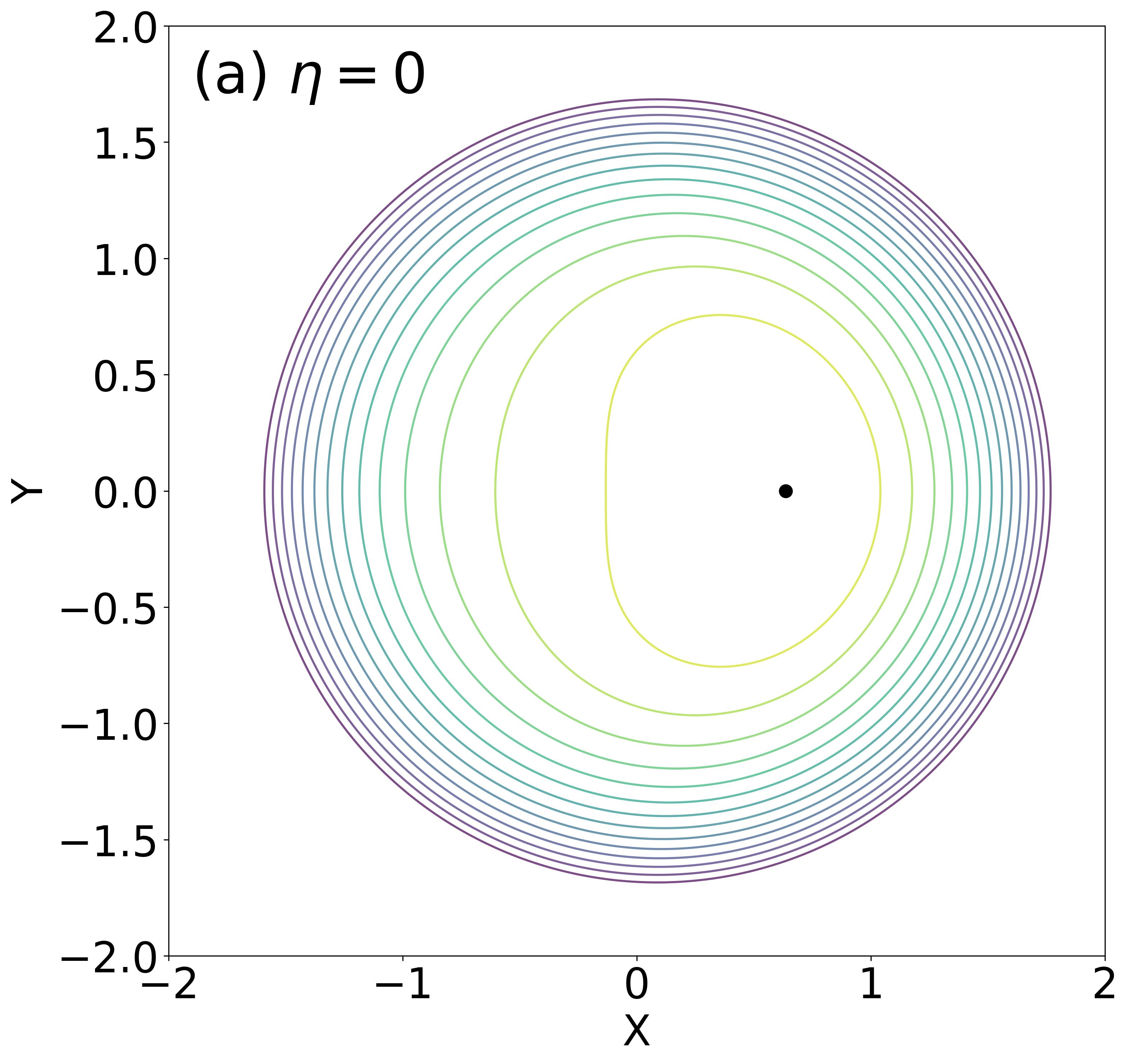} \label{eta_0}
    } 
    \hspace{0.05\textwidth}
    \subfigure{
        \includegraphics[width=0.4\textwidth]{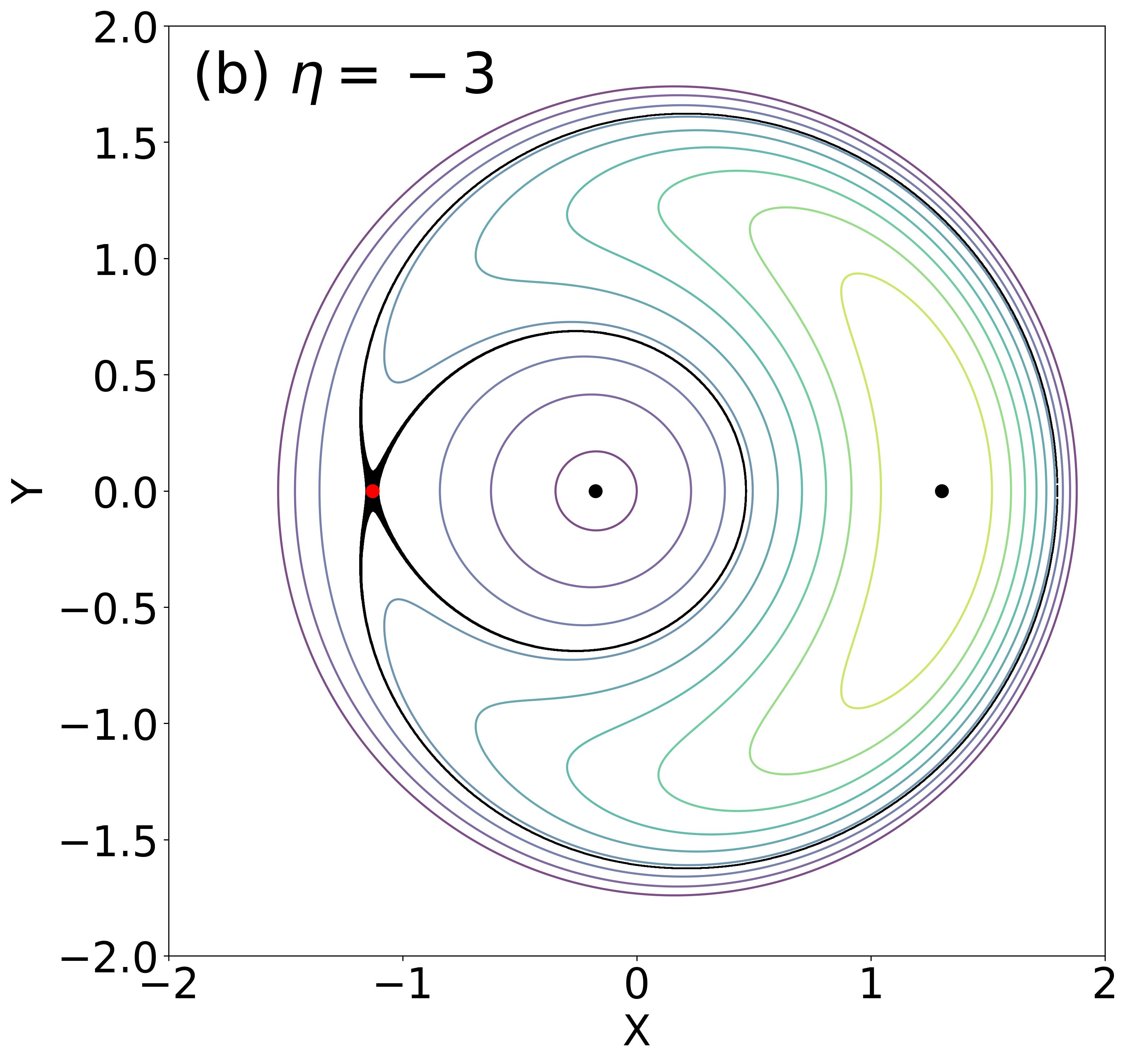} \label{eta_3}
    } 
    \caption{Level curves of the Hamiltonian (Eq. \ref{H}) for different $\eta$, plotted in the phase space of the conjugate variables  $X{=}\sqrt{\Phi}\cos{\varphi}$ and $Y{=}\sqrt{\Phi}\sin{\varphi}$. The black (red) dots mark the stable (unstable) fixed points, the thick black line in panel (b) marks the separatrix. Each curve represents the conserved Hamiltonian in the absence of dissipitive disk forces.  The color of the lines represents the values of the Hamiltonian.
   }
    \label{fig:level_curve}
    \end{figure}
    
\subsubsection{Escape criterion}
\label{sec:esc}

When planets are in resonances and under the stable criterion the resonant quantities ($\phi$,$\Phi$,$\eta$) remain fixed with a small amplitude libration. Here we attempt to answer what occurs if the above stability criterion is not satisfied. This state is commonly described as overstability \citep{2014AJ....147...32G, 2015ApJ...810..119D, 2017MNRAS.468.3223X}. In this case, the perturbations grow exponentially over time due to non-adiabatic effects (eccentricity damping in this case). Whether the overstability leads planets to escape directly or still stay in resonance can be further analyzed by the phase space topology.
    
Figure \ref{fig:level_curve} shows the Hamiltonian curves for different $\eta$ values in the phase space of the conjugate variables $X{=}\sqrt{\Phi}\cos{\varphi}$ and $Y{=}\sqrt{\Phi}\sin{\varphi}$. We note that $\eta$ represents the deviation from the exact resonance and the distance to the origin corresponds to the excited eccentricity ($\sqrt{X^2 + Y^2} {=} \sqrt{\Phi}{\sim} e$). 
   
Importantly, the value of $\eta$ sets the number of fixed points, and therefore determines the subsequent dynamical evolution. The fixed points can be obtained from $\partial H / \partial \varphi{=}\partial H/\partial \Phi {=}0$. It can be shown that \citep{Ferraz-Mello2007,2015MNRAS.451.2589B, Petit2017}, for $\eta {>} -3/2$, there is only one stable fixed point (black dot in Figure \ref{eta_0}), whereas for $\eta {<} -3/2$, there are two stable fixed points and one unstable point (red dot in Figure \ref{eta_3}).

 In overstable circumstances, the perturbations near the equilibrium grow over time. When the equilibrium value $\eta_{\rm eq} {>} -3/2$, the system's Hamiltonian curve expands as eccentricities are excited during resonant capture. However, the trajectory still bounds to the only fixed point. As a result, the system's eccentricity, period ratio, and resonant angle librate around their equilibrium values. Following \cite{2014AJ....147...32G} and \cite{2017MNRAS.468.3223X}, we refer to this state as an overstable trap --  although the equilibrium state is overstable and perturbations grow over time ($\delta \Phi \propto e^{\lambda_{\rm r}t}$, $\lambda_{\rm r}{>}0$), the system remains trapped with the resonant angle librating at a finite amplitude. This state is also termed as the limit cycle in \cite{2015ApJ...810..119D, 2022ApJ...925...38N, 2024A&A...686A.277A}. One example of an overstable trap can be found in Figure \ref{fig:overstab}.

\begin{figure}
     \centering
     \includegraphics[width=0.5\textwidth]{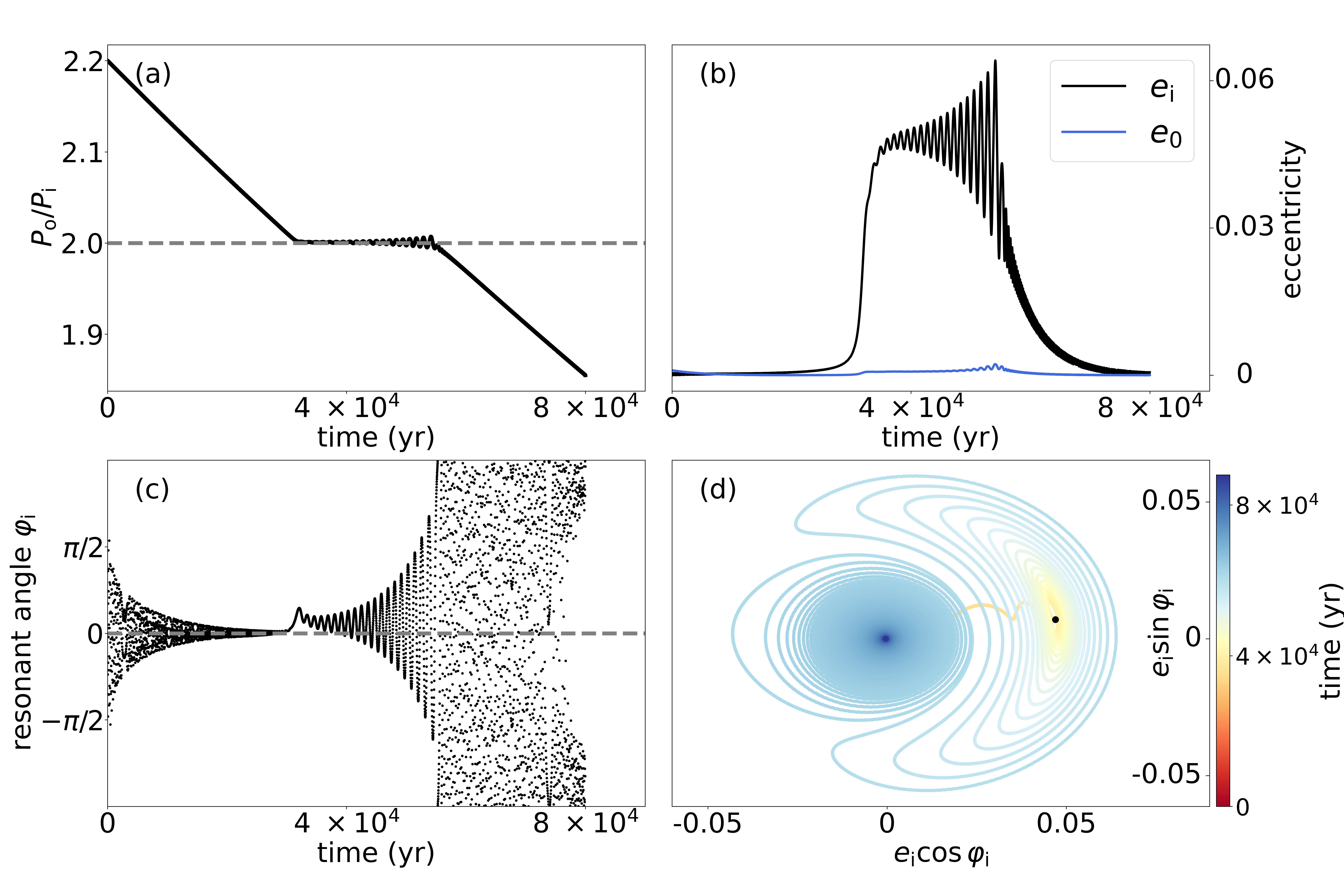}
     \caption{Escape from resonance in a two-planet system with convergent migration. 
     The planet and disk parameters are  $m_{\rm i}{=}1\ M_{\oplus}$, $m_{\rm o}{=}10\ M_{\oplus}$, $\tau_{\rm m} {= }5 \times 10^5 $ yr and $\tau_{\rm m} / \tau_{e} {= }200$.
 The system is temporarily trapped in a $2$:$1$ MMR during $t{\sim}2{-3}\times 10^{4}$ yr, and the inner planet reaches the equilibrium eccentricity with growing amplitude. The system escapes from resonance afterward, with $e_{\rm i}$ quickly damping to a very low value. Similarly,  $\varphi_{\rm i}$ features an increasing libration amplitude during the trap and eventually oscillates from $0$ to $2 \pi$. In phase space, the trajectory initially librates around the right fixed point (black dot).  As the amplitude grows, it ultimately escapes from the fixed point and settles into the origin. See \url{https://github.com/llh-astro/resonance/raw/main/gif/escape.gif}
}
     \label{fig:esc}
\end{figure}

   The separatrix emerges when $\eta_{\rm eq} {<} -3/2$ and the topology completely changes, as shown in Figure \ref{eta_3}. In this case, the system's Hamiltonian curve originally sits near the right stable fixed point. As the curve expands, it crosses the separatrix and enters the circulation zone centered around the left stable fixed point (from the right to the left black dot in Figure \ref{eta_3}). After that, the curve continuously shrinks towards the left black fixed point. As such, the eccentricity decreases, and the weaker resonant interaction cannot keep the planet pair in resonance anymore. The planets' period ratio departs from the exact commensurability, and the system eventually escapes from the resonance. One example of escape is shown in Figure \ref{fig:esc}.

   Based on Eq. \eqref{eq_eta}, $\eta_{\rm eq}{<}-3/2$ requires 
      $\Phi_{\rm eq} {= }(j-1)\beta\tau_{e}/(2j\tau_{\rm m}) {>} 1$,
   which is corresponding to 
    \begin{equation}
      \frac{\tau_{\rm m}}{\tau_{e}} < \left(\frac{3}{\mu_{\rm o}}\right)^{2/3}\frac{1}{4j}\left[\frac{(j-1)^{2}}{f_{\rm d,i}\alpha}\right]^{2/3} 
      \sim \mu_{\rm o}^{-2/3}. \label{esc_crit}
    \end{equation}
   The above escape criterion agrees with Eq. (28) of \cite{2014AJ....147...32G}.
   
   It is worth noting that this escape condition requires the system already to be in the overstable regime. In the case where the mass of the outer planet is much lower than that of the inner one ($q \gg 1$), the resonance is always stable. As such, the escape does not occur even when $\eta_{\rm eq} {<} -3/2$.

   To summarise, resonance trapping is always stable when the mass of the inner planet is much higher than that of the outer one ($q {\gg} 1$). However, when the mass of the outer planet is much higher than that of the inner one ($q {\ll} 1$), the outcome can be classified into three configurations: stable trap,  overstable trap, or escape. We further extend the analysis to arbitrary mass ratios $q$ and derive the corresponding criteria as follows: 

\begin{equation}
 \begin{cases}
 {\displaystyle   \frac{\tau_{\rm m}}{\tau_{e}} > \left(\frac{3}{\mu_o}\right)^{2/3} h(q)\ \ \left[\frac{(j-1)}{f_{\rm d,i}\alpha} \right]^{2/3} f(q)} \\
   {\displaystyle    }  
     \hfill  [\mbox{stability criterion}],  \vspace{0.1cm}\\
 {\displaystyle   \frac{\tau_{\rm m}}{\tau_{e}}< \left(\frac{3}{\mu_o}\right)^{2/3} \frac{h(q)}{4}\ \ \left[ \frac{(j-1)^{2}+j^{2}q}{f_{\rm d,i}\alpha + f_{\rm d,o}^{'}q^{2}} \right]^{2/3} }\\
   {\displaystyle   } 
        \hfill  [\mbox{escape criterion}],
\end{cases}
\label{eq:criteria}
\end{equation}
where  
\begin{eqnarray}
    h(q) & = & \frac{1}{(1+q\sqrt{\alpha})} 
               \frac{f_{\rm d,i}^{2}+f_{\rm d,o}^{'2}q^{2}\alpha}{j f_{\rm d,i}^{2}+(j-1)f_{\rm d,o}^{'2}q\sqrt{\alpha}},\\
    f(q) & = & 1-\left(\frac{f_{\rm d,o}^{'}}{f_{\rm d,i}}\right)^{2}q^{2}\alpha.
\end{eqnarray}
We note that $h(q)$ and $f(q)$ can also be written as   $h(q) {= }\tau_{\rm m}/\tau_{e}(e_{\rm i,eq}^{2}+e_{\rm o,eq}^{2})$, and $f(q){ =} 1-(e_{\rm o,eq}/e_{\rm i,eq})^{2}$. In the limit $q {\ll} 1$, Eq. \eqref{eq:criteria} is equivalent to Eqs. \eqref{stab_crit} and \eqref{esc_crit} . We show the detailed derivation in Appendix B.


\begin{figure*} 
    \centering
    \subfigure{
        \includegraphics[width=0.45\textwidth]{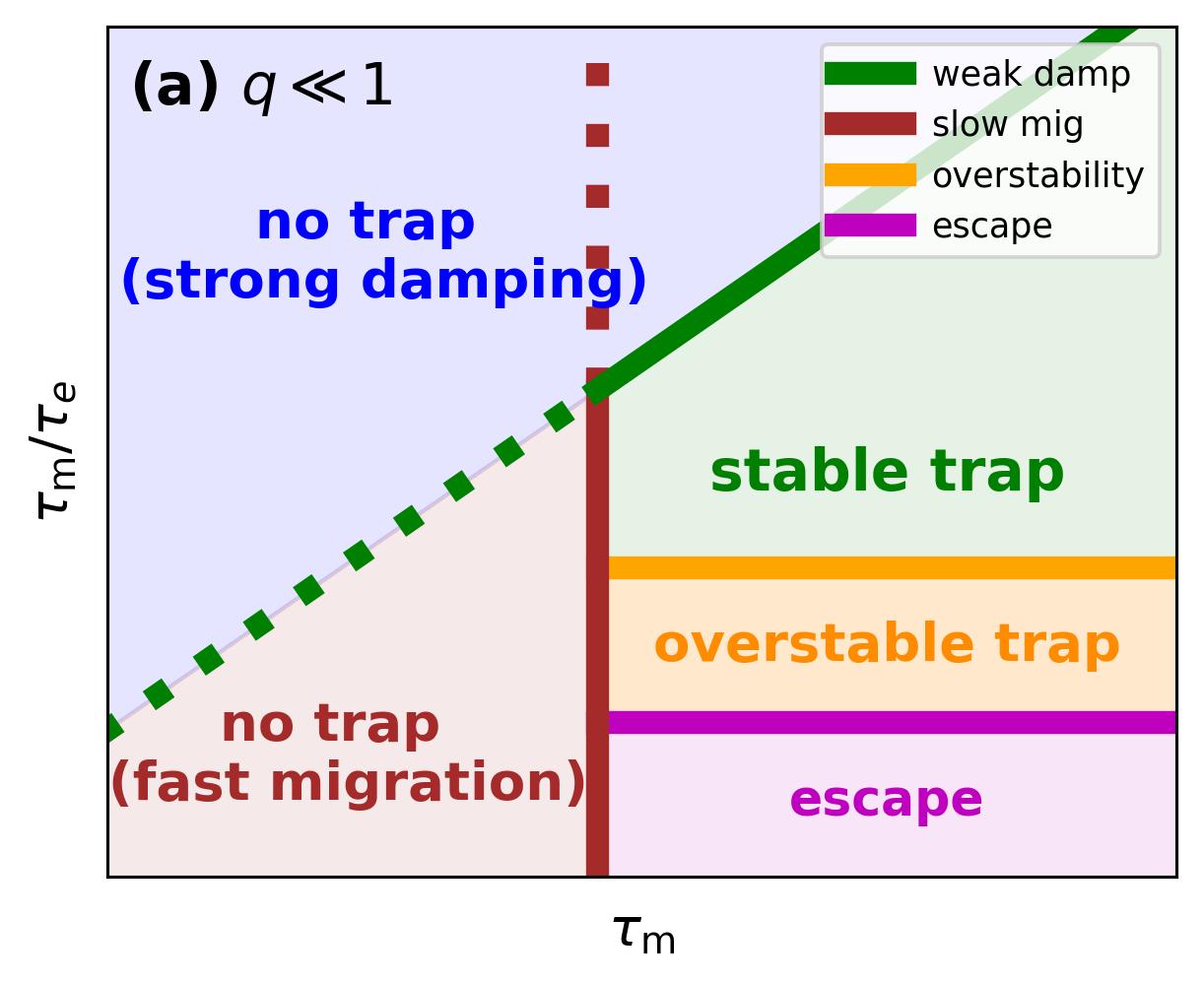} \label{fig:q_ll_1}
    }
    \hspace{0.05\textwidth}
    \subfigure{
        \includegraphics[width=0.45\textwidth]{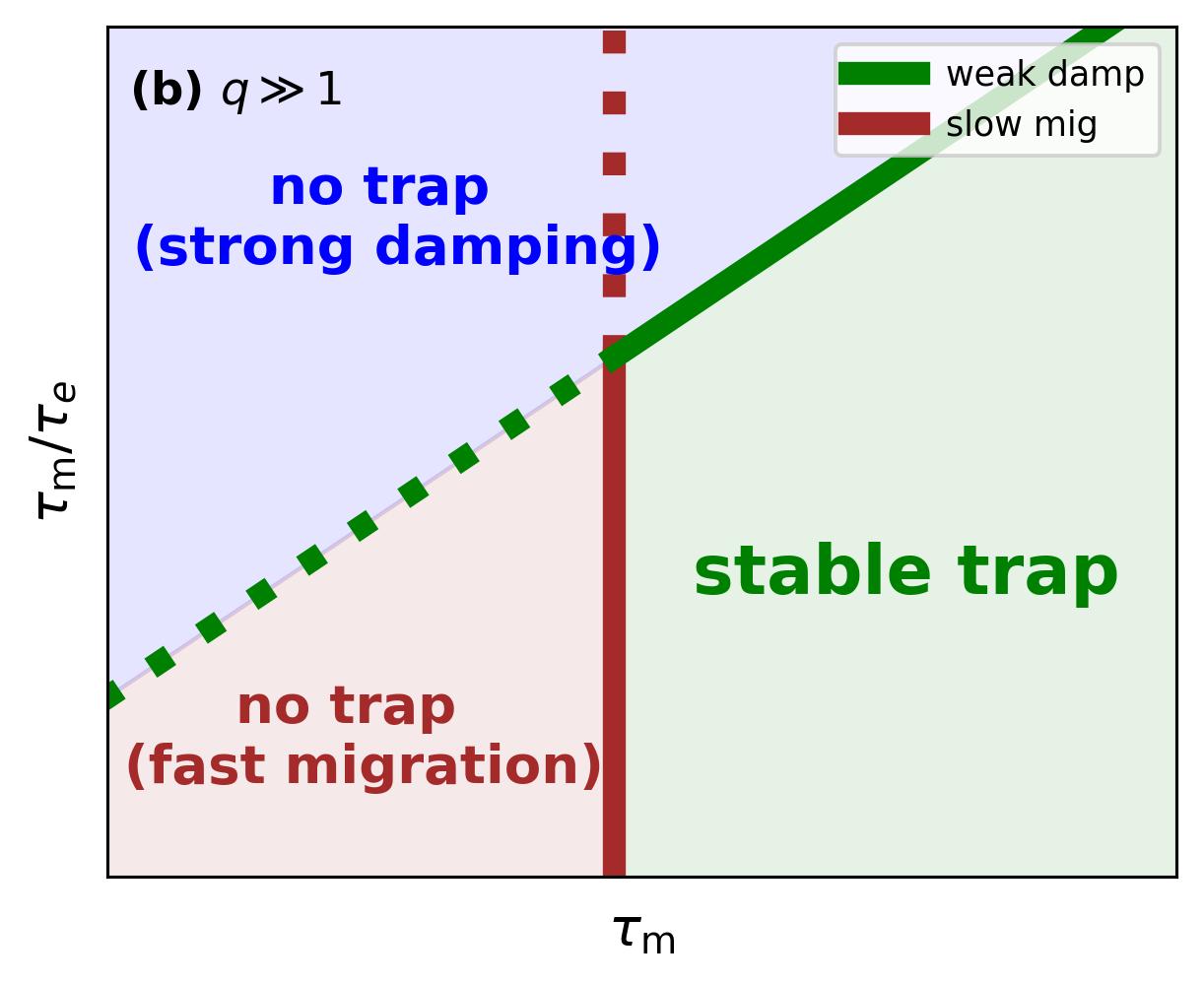} \label{fig:q_gg_1}
    }
    \caption{Analytical predictions of two-planet systems undergoing  convergent migration near the $2$:$1$ MMR in the limits of $q{\ll}1$ (left) and $q{\gg}1$ (right).  Blue region: planets directly bypass the MMR due to strong eccentricity damping. Brown region:  convergent migration is too fast to allow resonance capture. Green region:  planets get trapped in a stable resonance. Yellow region: despite overstability occurs, planets  still get trapped in the resonance, with libration of finite amplitude around the equilibrium point. Purple region: after a temporary trap, planets eventually escape resonance and continue the inward migration. The illustrative examples of the above regimes can be found in Figure $1{-}5$.} 
    \label{fig:illustration}
\end{figure*}

\subsection{A unified picture}
\label{sec:picture}

     We summarize the resonance capture outcomes for planet pairs that undergo convergent migration. Representative simulation results from each regime are also presented. 
    
    \begin{enumerate}[\textbullet]
        \item \textbf{No trap} (Figure \ref{fig:res_not_trap}): The planets directly bypasses the resonance. After crossing the resonance, the planets' eccentricities drop and the resonant angles vary from $0$ to $2 \pi$. 
        
        \item \textbf{Stable trap} (Figure \ref{fig:stab_trap}): The planets get locked in resonance,  and the system's trajectory converges towards an equilibrium value in the phase space. The planets' period ratio, eccentricities, and resonant angles remain at equilibrium values after capture.
        
        \item \textbf{Overstable trap } (Figure \ref{fig:overstab}): 
        The planets are locked in resonance with librating resonant angles. The eccentricities and the resonance angles of planets initially converge to the equilibrium values and, after a while,  librate with increasing amplitudes. The period ratio of the planets remains constant after capture.  In the phase space, the system's trajectory initially approaches the fixed point, then transitions into oscillations before ultimately stabilizing into a limit cycle around that point.
        
        \item  \textbf{Escape }(Figure \ref{fig:esc}): 
        The planets are temporarily locked in resonance before eventually escaping. The period ratio, eccentricities, and resonance angle of the planet pair initially converge to the equilibrium values. Then the eccentricities and resonance angles start to oscillate with growing amplitudes. Eventually, the planets leave resonance, with a shift in resonant angles from libration to circulation. In phase space, the system's trajectory initially approaches the fixed point; however, the oscillation gradually amplifies until it crosses the separatrix. The trajectory ultimately stabilizes towards the coordinate origin.
    \end{enumerate}

We point out that several previous studies have explored this issue in depth \citep{2023MNRAS.522..828H, 2014AJ....147...32G, 2015ApJ...810..119D, 2015MNRAS.451.2589B, Pichierri2018, 2023ApJ...946L..11B}. Nevertheless, most studies focus on one aspect of these issues at a time. Here, we provide a unified framework for studying the dynamics of both resonance capture and subsequent stability. This analysis is applicable to planet pair systems with arbitrary planet mass ratios and at different first-order mean motion resonances. An illustration of this framework can be found in Figure \ref{fig:illustration}. The results are presented in a $\tau_{\rm m}/\tau_{e}$–$\tau_{\rm e}$ parameter space, which links all dynamical regimes to the two key parameters that govern planet-disk interactions: angular momentum and eccentricity damping. We describe below how the variations in $\tau_{\rm m}$ and $\tau_{\rm e}$ influence the outcome.

  If convergent migration occurs too rapidly, the planets directly bypass the resonance without reaching the equilibrium state. As migration slows down, the effect of eccentricity damping becomes significant. Strong e-damping has two effects. First, it causes a reduction in planet's eccentricity, weakening the resonant interaction, since the resonant term is tied to planet's eccentricity (\eqref{n_eq}). Furthermore, the semi-major axis also decays due to orbital circularization. If the magnitude of such orbital decay term surpasses the resonant term, the planets will no longer remain in resonance. In this regard, planets cannot be captured when $\tau_{e}$ and/or $\tau_{\rm m}$ are short. No trap occurs in the upper and left parts of Figure \ref{fig:illustration}.

    When planets are trapped in resonance, the long-term stability of the system is related to the degree of eccentricity excitation. The damping effect ($\tau_{e}$) reduces the planet's eccentricity, making the system stable. On the other hand, convergent migration ($\tau_{\rm m}$) brings planets closer together, in turn exciting their eccentricities. Consequently, the relative eccentricity damping $\tau_{\rm m}/\tau_{e}$ is the key parameter in determining the stability. As can be seen from the top to the bottom of Figure \ref{fig:illustration}, the system tends to be more unstable as $\tau_{\rm m}/\tau_{e}$ decreases.  
    
    For the small inner planet case shown in Figure \ref{fig:q_ll_1}, if the relative damping $\tau_{\rm m}/\tau_{e}$ is sufficiently strong, the resonant trapping is statically stable, and the final state of the system stays at a fixed point in the phase space (stable trap regime). As the relative damping weakens (Eq. \ref{stab_crit}), the fixed point becomes unstable and the trajectory of the system begins to oscillate around the fixed point. It ultimately settles into a stable limit cycle (overstable trap).  The limit cycle expands as $\tau_{\rm m}/\tau_{e}$ continues to decrease. When the trajectory crosses the separatrix (Eq. \ref{esc_crit}), the system becomes unstable. In this regime, the oscillations grow over time and planets eventually leave resonance (escape). In contrast, for the case of a small outer planet shown in Figure \ref{fig:q_gg_1}, all traps are stable.

\begin{figure}
    \centering
    \includegraphics[width=0.99\linewidth]{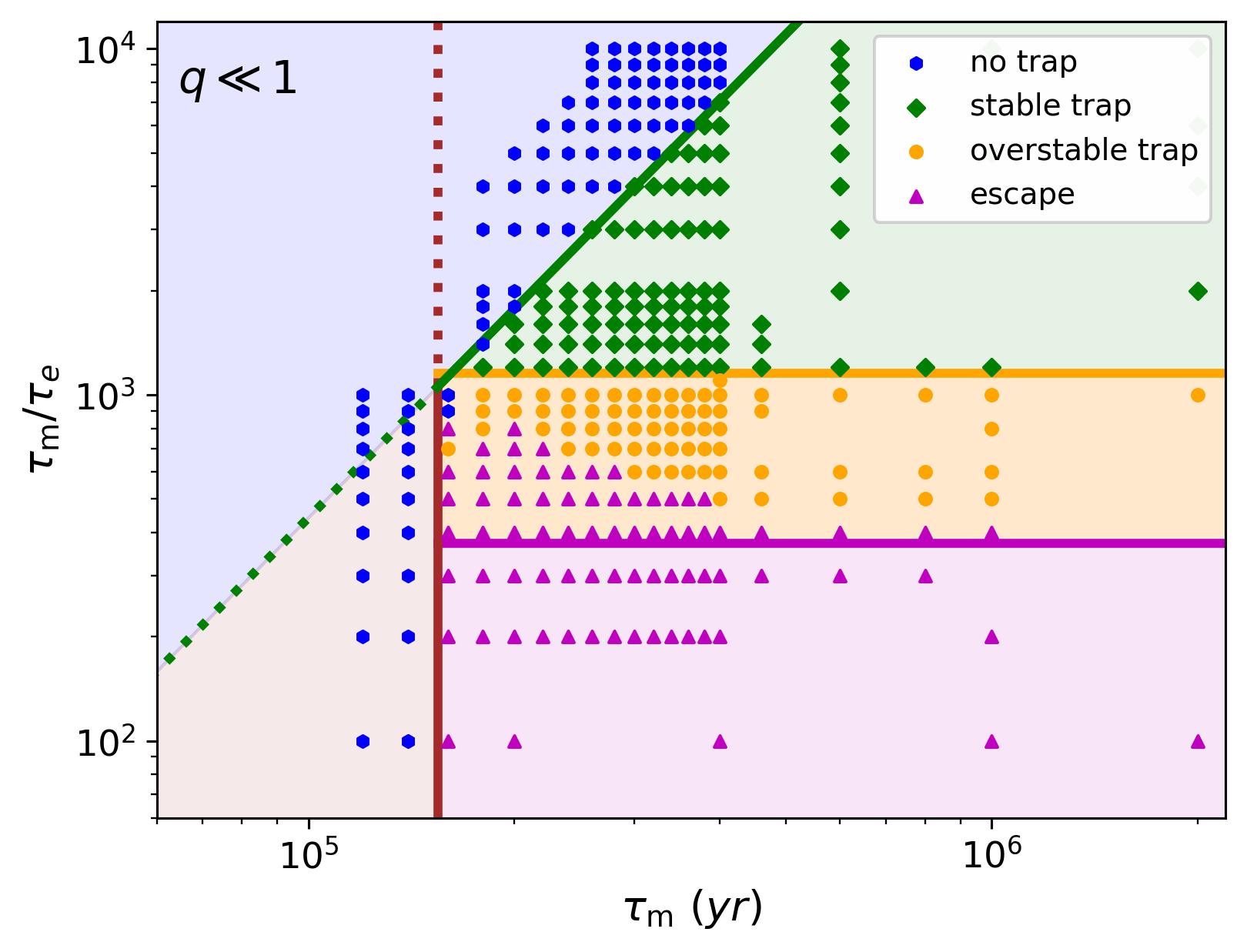}
    \caption{Simulations of the $2$:$1$ MMR capture and stability for two-planet systems with an inner to outer planet mass ratio $q$ of $0.1$. Blue: planets bypass resonance  due to rapid migration or strong eccentricity damping. Green: stable resonance capture. Yellow: overstable resonant capture. Purple: Resonance escape after temporary capture. 
    The simulations are consistent with the analysis outlined in Figure \ref{fig:q_ll_1}.  
    }
    \label{fig:q=1e-1}

    \vspace{1cm}

    \includegraphics[width=0.99\linewidth]{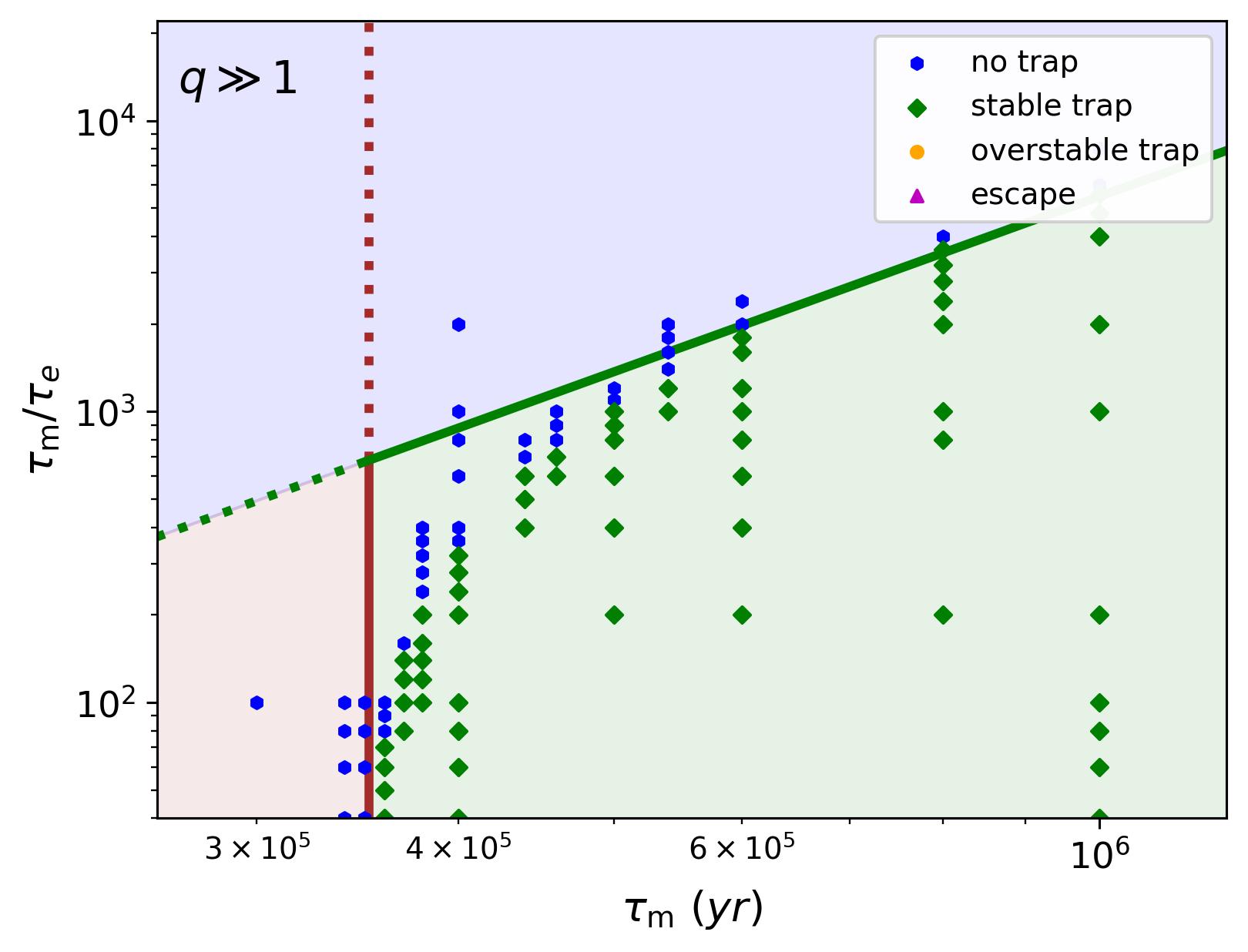}
    \caption{Same as Figure 7 but with an inner to outer planet mass ratio $q$ of $10$.
    }
    \label{fig:q=10}
\end{figure}

\begin{figure}
    \centering
    \includegraphics[width=0.99\linewidth]{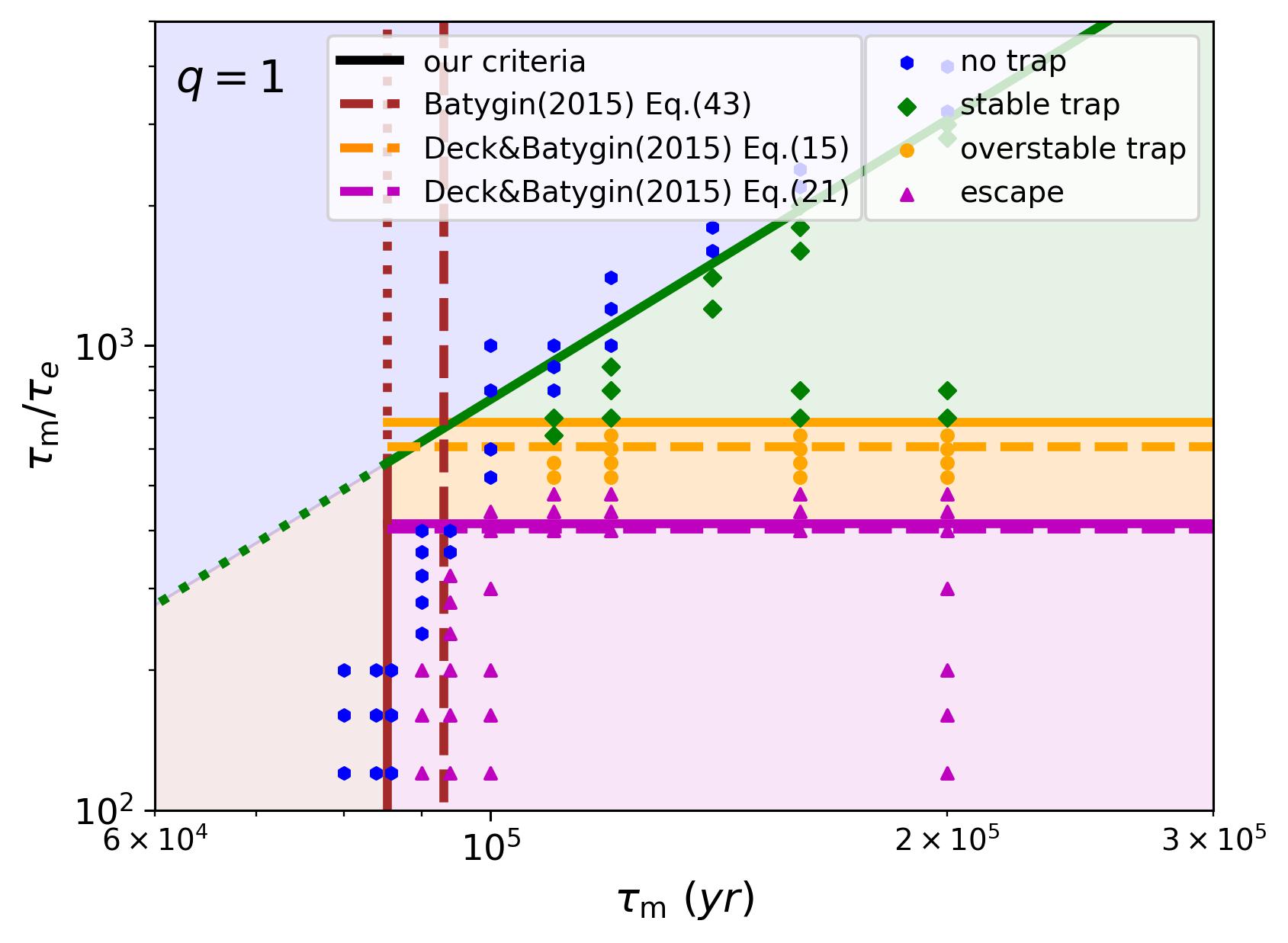}
    \caption{
    Same as Figure 7 but with equal-mass of $q{=}1$. The solid lines are the criteria derived in this study, whereas the dashed lines represent the trapping criteria of \cite{2015MNRAS.451.2589B} (their Eq. 43), and the stability and escape criteria of \cite{2015ApJ...810..119D} (their Eqs. 15 and 21).}
    \label{fig:q=1}
\end{figure}

\subsection{Resonance crossing through divergent migration}
\label{sec:div_mig}
 We have shown in the previous subsections that the convergent planet migration coupled with eccentricity damping can result in resonance trapping.  Literature theoretical work suggested that planets directly cross the resonance when the direction of their migration is divergent due to separatrix crossing \citep{Henrard1983,1999ssd..book.....M, 2024AJ....167..112W}. Here we also explain that divergent migration cannot lead to resonance capture in our framework.
 
 When the planet pair undergoes the divergent migration, the equilibrium state $\dot{n_{\rm i}}/n_{\rm i}  {=} \dot{n_{\rm o}}/n_{\rm o}$ (if it exists) can be written as
\begin{equation}
\begin{aligned}
   &  -[ (j-1) \mu_{\rm o} n_{\rm i} \alpha
    + j \mu_{\rm i} n_{\rm o} ]
    ( e_{\rm i} f_{\rm d,i} \sin{\varphi_{\rm i}}
    + e_{\rm o} f_{\rm d,o}^{'} \sin{\varphi_{\rm o}} )\\
    &=  -\frac{1}{\tau_{\rm m}}  + \frac{e_{\rm o}^2 
    - e_{\rm i}^{2}}{\tau_e},
    \label{n_eq_diver}
\end{aligned}
\end{equation}
where the relative orbital decay term ($ -\tau_{\rm m}^{-1}$) is negative compared to Eq. \eqref{n_eq}.
In order to derive equilibrium eccentricity, we repeat the procedure outlined in Sec. \ref{sec:criteria} by substituting $\dot{e_{\rm i}}{=}\dot{e_{\rm o}}{=}0$ into Eqs. \eqref{dot_ei} and \eqref{dot_eo} to eliminate $\sin{\varphi_{\rm i,o}}$ and  $\dot{\varpi_{\rm i}} {=}\dot{\varpi_{\rm o}} $ into Eqs. \eqref{dot_wi} and \eqref{dot_wo} to eliminate either $e_{\rm i}$ or $e_{\rm o}$. Then Eq. \eqref{n_eq_diver} reads
\begin{equation}
   \frac{e_{\rm i}^{2}}{\tau_{\rm e}} (1+q\sqrt{\alpha}) [j+ (j-1)\frac{f_{\rm d,o}^{'2}}{f_{\rm d,i}^{2}} q \sqrt{\alpha}]
    =  -\frac{1}{\tau_{\rm m}}.
\end{equation}
Noticeably, as long as $\tau_{\rm m}$ and $\tau_{\rm e}$ have the same sign, the left-hand side of this equation has the opposite sign as the right-hand side, and therefore the equality  never holds. This indicates that there is no equilibrium solution in the divergent migration circumstance, and even a small amplitude of divergent migration is sufficient for planets to cross the resonance.

\section{Simulations}
\label{sec:simulation}

In this section, we conduct numerical simulations to verify the aforementioned analytical criteria.  The numerical setup is described in Section \ref{sec:code}. We first test the $2$:$1$ resonance for planets with extreme mass ratios in Section \ref{sec:illustration}, and then extend the analysis for planets with arbitrary mass ratios in Section \ref{sec:compare}.  Other first-order resonances are investigated in Section \ref{sec:j_vary}.  

\subsection{Numerical setup}  
\label{sec:code} 

  We use the open-source N-body code \texttt{REBOUND} with the \texttt{WHfast} integrator \citep{2012A&A...537A.128R}. In simulations we also consider planets in coplanar orbits. The accelerations for angular momentum and eccentricity damping are given by \citep{Papaloizou2000, Cresswell2006, 2008A&A...482..677C, Pichierri2024} 
    \begin{eqnarray}
       {a_{\rm m}}  =  - \frac{\textbf{v}_{\rm pl}}{\tau_{\rm m}},  \ \ 
       {a_{\rm e}} =  -2 \frac{(\textbf{v}_{\rm pl} \cdot \textbf{r}_{\rm pl}) \cdot \textbf{r}_{\rm pl}}{r_{\rm pl}^2 \tau_{e}},
    \end{eqnarray}
   which are implemented through \texttt{REBOUNDx} \citep{2020MNRAS.491.2885T}.
    
   Initially, the inner planet is placed at $1$ au and the orbit of the outer planet is slightly away from the desired first-order resonance ( e.g., at $1.7$ au for the $2$:$1$ resonance). Both planets have circular orbits with random phase angles. To mimic convergent migration, only the outer planet undergoes inward migration on a timescale of $\tau_{\rm m}$, and the eccentricities of both planets are damped on a timescale of $\tau_{e}$. Simulations are terminated at $t {=} 2\tau_{\rm m}$, and the outcomes can be classified in different regimes, as described in Section \ref{sec:picture}.
    
    We consider three cases for the masses of the planet-pair: small inner planet ($m_{\rm i} {=} 1 \ M_{\oplus}$, $ m_{\rm o} {=} 10 \ M_{\oplus}$,  $q {=} 0.1$), small outer planet ($m_{\rm i} {=} 10\ M_{\oplus}$, $m_{\rm o} {=} 1 \ M_{\oplus}$, $q {=} 10$), and equal-mass planets ($m_{\rm i} {=}m_{\rm o} {=} 10\ M_{\oplus}$, $q {=} 1$).  We vary $\tau_{\rm m}$ and $\tau_{e}$ in each simulation. 

\subsection{Systems at two extreme mass ratios}
\label{sec:illustration}

Figure \ref{fig:q=1e-1} shows the simulations for the planet pair with the inner planet mass ten times lower than that of the outer one ($m_{\rm i} {=} 1 \ M_{\oplus}$, $ m_{\rm o} {=} 10 \ M_{\oplus}$,  $q {=} 0.1$), the results of which can be compared with the schematic presented in Figure \ref{fig:q_ll_1}.

We find that the simulations agree pretty well with the analytical estimates. No planets lead the resonant trap (blue dots) in the upper left part of Figure \ref{fig:q=1e-1}, when the weak damping (green line) and slow migration (brown line) criteria break. When those trapping criteria are satisfied and as $\tau_{\rm m} / \tau_{e}$ decreases, the simulations depict a transition from the stable trap (green dots) to overstable trap (yellow dots), and eventually enter escape (purple dots) regimes. These results are indeed separated by the two analytical criteria proposed in Sect. \ref{sec:analysis} (yellow and purple lines). 

The simulations do not fully match the analysis at $\tau_{\rm m} {\sim} 2\ \times 10^5 \ {\rm yr}$ and $\tau_{\rm m} / \tau_e {\sim} 6 \times 10^2$ in Figure \ref{fig:q=1e-1}. The theory predicts the planets in an overstable trap regime whereas in simulations planets actually leave the resonance. We hypothesize that the increased relative migration rate leads to more intense and nonlinear perturbations. This results in greater oscillation amplitudes around the equilibrium point. As such,  the system has a higher probability of crossing the separatrix and escaping resonance.

 Figure \ref{fig:q=10} demonstrates simulations for the planet pair with the inner planet mass ten times higher than that of the outer one ($m_{\rm i} {=} 10\ M_{\oplus}$, $m_{\rm o} {=} 1 \ M_{\oplus}$, $q {=} 10$), offering a comparison to the schematic illustrated in Figure \ref{fig:q_gg_1}. 
Again, no trap (blue dots) occurs in the upper and left parts of Figure \ref{fig:q=10}, when the weak damping and slow migration criteria are not satisfied. When these criteria are fulfilled, all traps correspond to stable traps (green dots).

We note that the no-trap regime occupies a slightly larger parameter space than theoretically expected. For instance, at $\tau_{\rm m} {\sim} 4 {\times} 10^{5} \ {\rm yr}$ and $\tau_{\rm m} / \tau_{e} {\sim} 4 \times 10^{2}$, while the trapping criteria are fulfilled, the planets are not in resonance.  This discrepancy arises from the simplification in deriving our slow migration criterion, which neglects the eccentricity damping effect (Eq. \ref{slow_crit}). When the eccentricity damping is incorporated, the interaction between the planets is further suppressed—owing to reduced eccentricities in the left-hand side of Eq. \ref{n_eq}—thereby prolonging the migration timescale on the right-hand side of Eq. \ref{n_eq}. Consequently, planets in reality bypass the resonance with slower migration rates compared to the simplified estimation from Eq. \ref{slow_crit}. 

Overall, we find that the analytical criteria align well with simulations in the circumstances of the planet pairs with extreme mass ratios.
\subsection{systems at arbitrary mass ratios}
\label{sec:compare}

The resonance dynamics has been extensively investigated in literature studies. \cite{2015MNRAS.451.2589B} proposed an adiabatic capture criterion for systems without eccentricity damping, linking the resonant libration time to the migration time. Subsequently, \cite{2023MNRAS.522..828H} derived a weak eccentricity damping criterion for planet pairs with extreme mass ratios ($q {\ll} 1$), while \cite{2023ApJ...946L..11B} formulated a generalized criterion applicable to arbitrary mass ratios. Regarding the post-capture evolution, \cite{2014AJ....147...32G} first proposed the concept of overstability and derived the corresponding analytic criteria for restricted three-body systems. \cite{2015ApJ...810..119D} further extended the analysis for planet pairs with arbitrary mass ratios.

In this subsection, we conduct numerical simulations to validate our theoretical criteria and compare our results with  those from literature studies. The detailed derivation of the $q$-dependent overstability and escape criteria are provided in Appendix B. As a point of comparison, our weak damping criterion is equivalent to that obtained by \citet{2023ApJ...946L..11B}, while our other criteria exhibit the same $\mu$-scaling as those in previous works but differ in their numerical prefactors: $(\tau_{\rm m})_{\rm slow,mig} {\sim} \mu^{-4/3}$, $(\tau_{\rm m}/\tau_{e})_{\rm overstab},(\tau_{\rm m}/\tau_{e})_{\rm esc} {\sim} \mu^{-2/3}$. Therefore, we compare our slow migration criterion with the adiabatic criterion in \cite{2015MNRAS.451.2589B} and our stability and escape criteria with those in \cite{2015ApJ...810..119D}. Additionally, we perform simulations to investigate the $q$-dependence of our overstability and escape criteria, as detailed in Appendix B. 

Figure \ref{fig:q=1} illustrates the simulations for the planet pair of equal-mass ($m_{\rm i} {=} m_{\rm o} {=} 10\ M_{\oplus}$, $q {=} 1$). The colored symbol depicts the simulation results, the solid lines represent our analytical criteria (Eq. \ref{eq:criteria}), and the dashed lines correspond to the criteria from \cite{2015MNRAS.451.2589B} (their Eq 43) and \cite{2015ApJ...810..119D} (their Eqs 15 and 21). As can be seen, both the criteria derived in this work and those from previous studies match the numerical results relatively well. A more detailed comparison that accounts for the dependence on the mass ratio $q$ is presented in Appendix B.

\subsection{Systems at arbitrary j:j-1 MMR}
\label{sec:j_vary}

In previous sections we focus on the circumstances of the $2$:$1$ resonance as a representative case. Since our analytical framework can be applied to other first-order  $j$:$j-1$ MMRs,  we also perform such numerical verification as follows.

\begin{figure}
    \centering
    \includegraphics[width=1\linewidth]{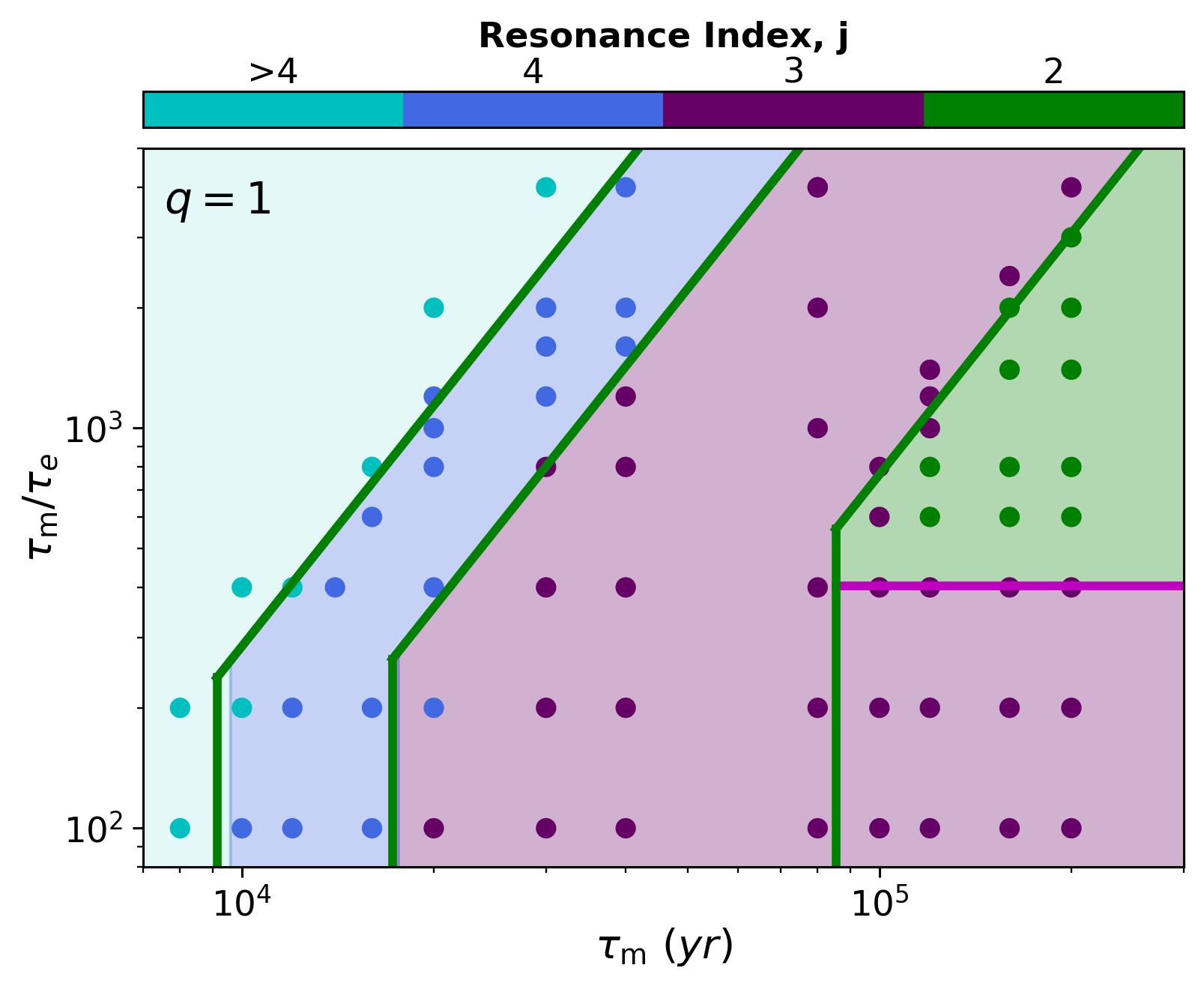}
    \caption{
    Simulations of  $j$:$j-1$ MMR capture and stability for two-planet systems with equal masses. Color dots represent systems' final resonant states.
    Green lines show $j$:$j-1$  trapping criteria, while purple indicates the $2$:$1$ escape criterion.}
    \label{fig:j_vary}
\end{figure}

Figure \ref{fig:j_vary} shows the resonant states in which the systems finally settle for equal-mass planet pairs ($m_{\rm i} {=}m_{\rm o} {=} 10\ M_{\oplus}$). The colored dots depict the numerical results, the green lines represent the trapping criteria, while the purple line represents the escape criterion. Note that for equal-mass planet pairs, our criteria predict that all traps are stable (no overstable trap and escape regimes) at $j {\geq} 3$ (Eq. \ref{eq:criteria}). Therefore, we only exhibit the escape criterion for the $2$:$1$ MMR in Figure \ref{fig:j_vary}.

Our criteria show good agreement with simulations. As the disk effects become stronger (e.g., $\tau_{\rm m}$ decreases and $\tau_{\rm m}/\tau_{e}$ increases), planets can bypass the lower $j$:$j-1$ MMRs and evolve into more compact orbits.

\section{Discussion}
\label{sec:discuss}
In this section, we outline a few limitations in our model.
   First, for simplicity we treat $\tau_{\rm m}$ and $\tau_{e}$ as free but constant parameters and two planets have the same eccentricity damping timescales. However, in reality $\tau_{\rm m}$ and $\tau_{e}$ depend on the gas disk mass, the planet mass, and orbital distance. For instance, in the linear type I regime, both $\tau_{\rm m}$ and $\tau_e {\propto}\mu^{-1}$, while the two trapping criteria exhibit steeper decreasing functions of $\mu$ such that $\tau_{\rm m,\ \text{slow\ mig}} {\propto} \mu^{-4/3}$, and $(\tau_{\rm m}\tau_{e})_{\rm \text{weak\ damp}} {\propto} \mu^{-2}$.  This suggests that planets become more easily trapped as their masses increase. Nevertheless, we note that the relation between planet mass and trapping outcome behavors more complicated. When the planets' masses further increase and exceed the gap-opening threshold, they gradually switch to Type-II migration \citep{Lin&papaloziou1986, Kanagawa2018}. In this regime, the planet's mass influences migration and eccentricity damping in a non-trivial way \citep{Kanagawa2018,Kajtazi2023, Pichierri2023, Pichierri2024}.  This additional complexity is not accounted for in our current analysis.
   On the other hand, since the planets can grow their masses at different rates, the planet mass ratio could change over time.  
   For instance, if the inner planet is initially more massive than the outer planet ($q {\gg }1$), the resonance trapping is in the stable regime. When the outer planet grows and the mass ratio decreases, the system may transition into an overstable trap or even escape.  Future studies should explore how these interactions modify the trapping criteria and how the evolving planet mass ratio affects the long-term dynamics of the system.

  Furthermore, both hydrodynamic simulations and analytical work have shown that the migration and eccentricity damping rates differ from those in the near circular case when planets attain high eccentricities \citep{2008A&A...482..677C, Bitsch2011, Bitsch2013, Xu&Lai2018, Ida2022, Pichierri2023, Pichierri2024, Fairbairn2024}. For instance, \cite{Xu&Lai2018} showed that nonlinear eccentricity damping can lead to higher equilibrium eccentricities, enhance the stability of captured resonances.

  In addition, our analysis carried out so far assumes a laminar disk. In a more realistic turbulent disk, turbulence-induced density fluctuations can generate additional stochastic torques \citep{Laughlin2004,Ogihara2006, rein2012, Okuzumi&Ormel2013, Huhn2021, Wu2024}. These effects are not accounted for in the aforementioned simplified framework. Such stochastic torques may considerably alter the resonance trapping outcomes \citep{Batygin&Adams2017, 2023ApJ...948...12G, Chen2025}. 

  Lastly, we only consider a static disk without considering the gas density decay due to disk dissipation. As disk gas disperses, the planet migration slows down and resonance trapping is easier to occur. However, for stability/overstability and escape criteria, the situation is less straightforward. This is because $\tau_{\rm m}/\tau_{e}$ depends on the aspect ratio of the disk but not the surface density. The disk thermal structure relies on the heating mechanism. While our framework remains applicable and can predict the evolution of resonance configurations as long as $\tau_m$
  and $\tau_e$
  are specified, understanding how these timescales change with disk dissipation is essential for making such predictions. However, this aspect is not within the scope of our current study.
   
\section{Conclusion}
\label{sec:conclusion}
We have constructed an analytical framework suited to study the resonant trapping, stability, and escape conditions for two-planet systems that undergo convergent type I migration in the protoplanetary disks.
The planet-disk interaction is assumed to be parameterized by two key quantities: angular momentum damping and eccentricity damping timescales $\tau_{\rm m}$ and $\tau_{e}$.
We present the analytical criteria for general $j$:$j-1$ resonances and for planet pairs with arbitrary mass ratios. 
The outcomes can be classified into four regimes: direct resonance crossing, stable resonance trap, overstable resonance trap and escape after temporary trap.  The proposed criteria and regimes, which are validated through numerical N-body simulations, can be illustrated in a $\tau_{\rm m}/\tau_{\rm e}$-$\tau_{e}$ plot (Figure \ref{fig:illustration}).

 We find that the planets can be trapped in resonances under slow migration and weak damping (Eqs. \ref{slow_mig} and \ref{weak damping}). When the mass of the inner planet is comparable to or less massive than the outer planet  ($q {\lesssim} 1$), after capture, the stability weakens as $\tau_{\rm m}/\tau_{e}$ decreases. The system transitions from a stable trap to an overstable trap (Eq. \ref{stab_crit}), and finally to escape (Eq. \ref{esc_crit}). This corresponds to the resonant states changing from statically stable to dynamically stable, and ultimately to dynamically unstable.  
In contrast, when the mass of the inner planet is much more massive than the outer planet ($q {\gg} 1$), the traps remain permanently stable.
The key formulas of different regimes  are presented in Eqs.~\eqref{eq:trap_criteria} and \eqref{eq:criteria}. Future studies are needed to connect this theoretical framework with observations \citep{Dai2024,Hu2025}.


\begin{acknowledgements}
The authors appreciate the constructive comments from the referee.  BL is supported by the National Key R\&D Program of China (2024YFA1611803), the National Natural Science Foundation of China (Nos. 12222303 and 12173035), and the start-up grant of the Bairen program from Zhejiang University. The simulations and analysis presented in this article were carried out on the SilkRiver Supercomputer of Zhejiang University.
Software: \texttt{REBOUND} \citep{2012A&A...537A.128R} and \texttt{REBOUNDX} \citep{2020MNRAS.491.2885T}.
      
\end{acknowledgements}


\bibliographystyle{aa}
\bibliography{main.bib}


\appendix

\section*{Appendix A: The derivation of the excited eccentricity in the classical pendulum model}
\label{sec:appendix A}
\renewcommand{\theequation}{A\arabic{equation}}
\setcounter{equation}{0} 
When the planets are in resonance, the second time derivative of the resonant angle $\varphi_{\rm i}$ and $\varphi_{\rm o}$ can be written as
\begin{equation}
    \Ddot{\varphi_{\rm i}} {=} \Ddot{\varphi_{\rm o}} {=} j\dot{n_{\rm o}} + (1-j)\dot{n_{\rm i}}, \label{ddot_phi}
\end{equation}
where the contribution from $\dot{\varpi_{\rm i}}$ and $\dot{\varpi_{\rm o}}$ are neglected. With no additional effects from the gaseous disk such as angular momentum damping and eccentricity damping, Lagrange’s equations can be written as
\begin{subequations}
\begin{align}
     \dot{n_{\rm i}} &     =  -3(j-1) \mu_{\rm o} n_{\rm i}^{2} \alpha 
                          ( e_{\rm i}f_{\rm d,i}\sin{\varphi_{\rm i}} + e_{\rm o}f_{\rm d,o}^{'}\sin{\varphi_{\rm o}} ), \label{a}\\
     \dot{n_{\rm o}}    &   =  3j \mu_{\rm i} n_{\rm o}^{2} 
                          ( e_{\rm i}f_{\rm d,i}\sin{\varphi_{\rm i}} + e_{\rm o}f_{\rm d,o}^{'}\sin{\varphi_{\rm o}} ), \label{b}\\
     \dot{e_{\rm i}}      &  =  -\mu_{\rm o} n_{\rm i} \alpha f_{\rm d,i} \sin{\varphi_{\rm i}}, \label{c} \\ 
     \dot{e_{\rm o}}     &  =  -\mu_{\rm i} n_{\rm o} f_{\rm d,o}^{'} \sin{\varphi_{\rm o}}. \label{d} 
\end{align}
\end{subequations}
Substituting $\dot{n_{\rm o}}$ and $\dot{n_{\rm i}}$ in Eq. \eqref{ddot_phi} with Eqs. \eqref{a} and \eqref{b}, and combing Eqs. \eqref{c} and \eqref{d}, we obtain
\begin{eqnarray}
    \Ddot{\varphi} {=} 3 [(j-1)^{2}\mu_{\rm o}n_{\rm i}^{2}\alpha + j^{2}\mu_{\rm i}n_{\rm o}^{2}](e_{\rm i}f_{\rm d,i} - e_{\rm o}f_{\rm d,o}^{'}) \sin{\varphi}, \label{}\\
    \dot{e_{\rm i}}f_{\rm d,i} - \dot{e_{\rm o}}f_{\rm d,o}^{'} = -(\mu_{\rm o}n_{\rm i}\alpha f_{\rm d,i}^{2} + \mu_{\rm i}n_{\rm o} f_{\rm d,o}^{'2}) \sin{\varphi}, \label{ei_eo}
\end{eqnarray}
where we assume $\sin{\varphi} \equiv \sin{\varphi_{\rm i}} = -\sin{\varphi_{\rm o}} $. The equation for $\Ddot{\varphi}$ can be simplified to the form 
\begin{equation}
    \Ddot{\varphi} {=} -\omega_{0}^{2} \sin{\varphi},
\end{equation}
where we take 
\begin{equation}
    \omega_{0}^{2} = -3 [(j-1)^{2}\mu_{\rm o}n_{\rm i}^{2}\alpha + j^{2}\mu_{\rm i}n_{\rm o}^{2}](e_{\rm i}f_{\rm d,i} - e_{\rm o}f_{\rm d,o}^{'}).
\end{equation}
The solution can be described as a libration of $\varphi$. The total energy $E$ of the system is the sum of the kinetic energy and potential energy. Hence $E = \frac{1}{2}\dot{\varphi}^2 + 2\omega_{0}^{2}\sin^{2}{\varphi /2}$.
It is clear that the maximum libration occurs when $\dot{\varphi}=0$ at $\varphi=\pm \pi$. This implies
$E_{\rm max} = 2 \omega_{0}^{2}$. We now set the value of $E$ equal to $E_{\rm max}$ and consider $\dot{\varphi}$ given by
\begin{equation}
    \dot{\varphi} = \pm 2\omega_{0}\cos{\varphi/2}.
\end{equation}
We can relate the variation of $\dot{\varphi}$ to the variation in $e_{\rm i,o}$ by means of Eq. \eqref{ei_eo}. This gives
\begin{equation}
    d(e_{\rm i}f_{\rm d,i} - e_{\rm o}f_{d,o^{'}}) {=} \pm \frac{(\mu_{\rm o}n_{\rm i}\alpha f_{\rm d,i}^{2} + \mu_{\rm i}n_{\rm o} f_{\rm d,o}^{'2})}{\omega_{0}} \sin{\frac{\varphi}{2}}d{\varphi}. \label{e_phi}
\end{equation}
Substitute $\omega_{0}$ into Eq. \eqref{e_phi} and integrate it, we obtain
\begin{equation}
    e_{\rm i}f_{\rm d,i} - e_{\rm o}f_{\rm d,o}^{'} = 
    -\frac{3^{1/3}(\mu_{\rm o}n_{\rm i}\alpha f_{\rm d,i}^{2} + \mu_{\rm i}n_{\rm o}f_{\rm d,o}^{'2})^{2/3}}
    {[(j-1)^{2}\mu_{\rm o}n_{\rm i}^{2}\alpha + j^{2}\mu_{\rm i}n_{\rm o}^{2}]^{1/3}}\cos{\frac{\varphi}{2}}.
\end{equation}
The maximum eccentricity occurs when $\varphi = 0$, which provides the eccentricity described in Section \ref{sec:slow_mig}.

\section*{Appendix B: Extension of stability and escape criteria to arbitrary mass ratios}
\renewcommand{\theequation}{B\arabic{equation}}
\setcounter{equation}{0}
\subsection*{B.1: Escape criterion}
For the cases of $q \ll 1$ and $q \gg 1$, the system follows the same Hamiltonian equation: 
\begin{equation}
    -H=\eta \Phi + \Phi^{2} -\Phi^{1/2}\cos{\varphi},
\end{equation}
where
\begin{equation}
 \begin{cases}
 {\displaystyle   \Phi \sim \left(\frac{3}{\mu_{\rm o}}\right)^{2/3} [\frac{(j-1)^{2}}{f_{\rm d,i}\alpha}]^{2/3} \frac{e_{\rm i}^{2}}{4},  
    \ \ \hfill  [\mbox{$q \ll 1$}],  \vspace{0.1cm} }\\
 {\displaystyle   \Phi \sim \left(\frac{3}{\mu_{\rm i}}\right)^{2/3} [\frac{j^{2}}{f_{\rm d,o}^{'}}]^{2/3} \frac{e_{\rm o}^{2}}{4},
    \ \ \hfill  [\mbox{$q \gg 1$}]. }
\end{cases}
\end{equation}
Therefore, we hypothesize the existence of a form for $\Phi$ that satisfies the Hamiltonian equation and degenerates to Eq. \eqref{Phi} in the extreme mass ratio cases:
\begin{equation}
    \Phi = \left(\frac{3}{\mu_{\rm o}}\right)^{2/3} [\frac{(j-1)^{2} + j^{2}q}{f_{\rm d,i}\alpha + f_{\rm d,o}^{'}q^{2}}]^{2/3} \frac{e^{2}}{4}, \label{Phi}
\end{equation}
where $e^{2} = e_{\rm i}^{2} + e_{\rm o}^{2}$. When $\eta_{\rm eq} {<} -3/2$ (equivalently $\Phi_{\rm eq} > 1$),  a separatrix point occurs and the system can cross this separatrix and escape from resonance (see Section \ref{sec:esc}). By substituting $e_{\rm i,eq}$ and $e_{\rm o,eq}$ (Eq. \ref{e_eq}) into Eq. \eqref{Phi},  we obtain the general escape criterion as
\begin{equation}
    \frac{\tau_{\rm m}}{\tau_{e}} < \left(\frac{3}{\mu_o}\right)^{2/3} \frac{h(q)}{4}\ \ \left[ \frac{(j-1)^{2}+j^{2}q}{f_{\rm d,i}\alpha + f_{\rm d,o}^{'}q^{2}} \right]^{2/3}.
\end{equation}
where
\begin{eqnarray}
    h(q) & = & \tau_{\rm m}/\tau_e \left( e_{\rm i,eq}^{2} + e_{\rm o,eq}^{2}\right) \nonumber \\
    & = & \frac{1}{(1+q\sqrt{\alpha})} \frac{f_{\rm d,i}^{2}+f_{\rm d,o}^{'2}q^{2}\alpha}{j f_{\rm d,i}^{2}+(j-1)f_{\rm d,o}^{'2}q\sqrt{\alpha}}.
\end{eqnarray}

\subsection*{B.2: Stability criterion}
The stability of the system decreases with the mass ratio $q$. When the outer planet is comparable or more massive than the inner planet ($q \lesssim 1$), the system may become overstable (see Eq. \ref{stab_crit}). However, all traps remain stable when the inner planet is more massive. Therefore, we introduce a factor $f(q)$ to fit the general stability criterion, which is given by
\begin{equation}
    \frac{\tau_{\rm m}}{\tau_{e}} > \left(\frac{3}{\mu_o}\right)^{2/3} h(q)\ \ \left[\frac{(j-1)^{2}}{f_{\rm d,i}\alpha} \right]^{2/3} (j-1)^{2/3} f(q),
\end{equation}
where
\begin{equation}
    f(q) = 1-\frac{e_{\rm o,eq}^{2}}{e_{\rm i,eq}^{2}}\
     =  1-\left(\frac{f_{\rm d,o}^{'}}{f_{\rm d,i}}\right)^{2}q^{2}\alpha.
\end{equation}
When $q \ll 1$, $f(q) \sim 1$, the system is dominated by the outer planet. As $q$ increases,  $f(q)$ decreases, indicating that the system becomes more stable. When $e_{\rm o,eq} > e_{\rm i,eq}$, $f(q)<0$, indicating that the system is dominated by the inner planet and all traps are stable traps.

\subsection*{B.3: Dependence on planet mass ratio}
    We run simulations to test our proposed overstability and escape criteria with different planet mass ratios by fixing $m_{\rm i} {=} 10\ M_{\oplus}$. 
     We set $\tau_{\rm m} {=} 3\ (\tau_{\rm m})_{\rm slow, mig}$ to fulfill the trapping criteria and vary $\tau_e$.
    The result is presented in Figure \ref{fig:q_dep}, where the solid and dashed lines denote the analytical criteria from ours and from \citet{2015ApJ...810..119D}. 

    The simulations agree pretty well with the analytical expectation at $q {\leq} 1$, with our criteria slightly better than \citet{2015ApJ...810..119D}. The simulations depict a transition from stable trap (green squares) to overstable trap (yellow circles), and eventually become escape (purple triangles) as $\tau_{\rm m} / \tau_{e}$ decreases, and our analytical criteria separate these three regimes. However, at $q {\geq} 1$, there are still two mismatches between simulations and analysis in Figure \ref{fig:q_dep}. 
    
    First, the stability/overstability and escape criteria exhibit an opposite trend in q, with the two lines intersecting at $q {\sim} 2$. For $q {\gtrsim} 2$, one would expect that the systems in the region below the overstability criterion (yellow line) should escape. However, simulations show that all systems in this region are trapped overstably (no escape occurs). We attribute this to the fact that when $q {\gtrsim} 2$, the angular momentum deficit (AMD) \citep{1997A&A...317L..75L} of the inner planet is higher than that of the outer planet. The exchange of angular momentum between these two planets weakens the oscillations of the amplitude of the outer planet's eccentricity (see Appendix B.4 for detailed derivation). Thus, the perturbations cannot grow sufficiently to cause escape, and the system is refined in the overstable trap regime. The solid blue line in Figure \ref{fig:q_dep} represents the AMD criterion up to the second-order eccentricity terms. For comparison, the blue dotted line is the AMD criterion with only the first-order eccentricity terms.  We find that the the AMD criterion with the second-order eccentricity terms separates the overstable trap and escape regimes at $q{\gtrsim}2$, which is in good agreement with our simulations.    
    
    Second, our overstable criterion suggests at $q {\geq} 3$, all traps are stable. However, in Figure \ref{fig:q_dep}, it is evident that near $q \sim 4$ and $\tau_{\rm m} / \tau_{e} \sim 2 \times 10^{2}$, the planets remain in an overstable state. We suspect the reason might be that we only considered the first-order term of the disturbing function. When the second-order term is included, the angular momentum transfer from the inner planet to the outer one becomes weaker. This leads to the shift of the regimes between stable and overstable to a higher $q$. A detailed analytical investigation for the offset seen in Fig. B.1 is nevertheless beyond the scope of this work.

    Besides, we also extend the test to the $3$:$2$ MMR in Fig. \ref{fig:q_dep_3_2}. We also find that our analysis agrees well with the numerical simulations.

\renewcommand{\thefigure}{B.\arabic{figure}}
\setcounter{figure}{0} 
    \begin{figure}
     \centering
     \includegraphics[width=0.5\textwidth]{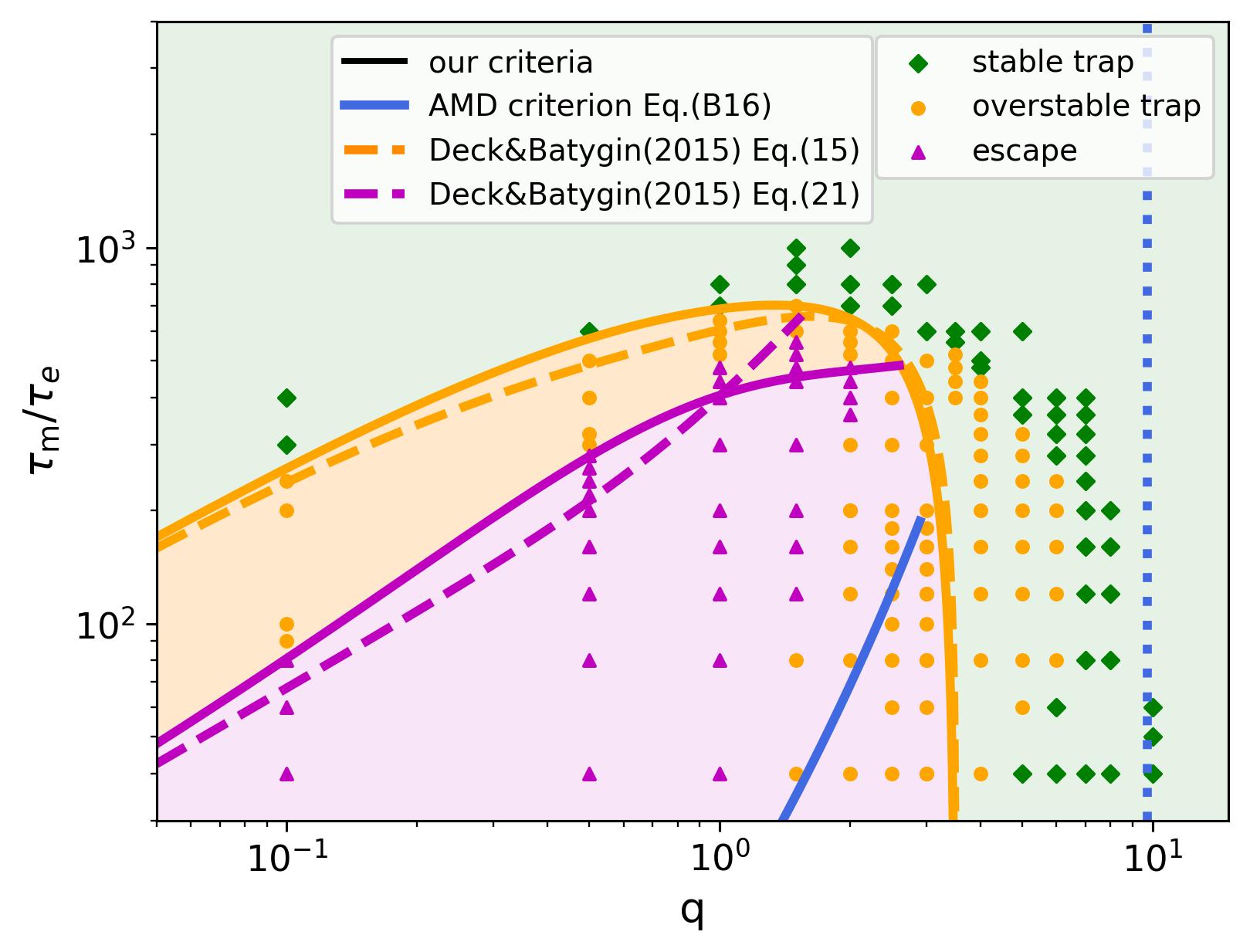}
     \caption{Simulations of $2$:$1$ MMR capture and stability in $q-\tau_{\rm m}/\tau_{e}$ parameter space. The mass of the inner planet $m_{\rm i} = 10\ M_{\oplus}$. Green: stable resonance capture. Yellow: overstable resonant capture. Purple: resonance escape after temporary capture.  The solid lines are the criteria derived in this study, whereas the dashed lines represent the stability and escape criteria of \cite{2015ApJ...810..119D} (their Eqs. 15 and 21). The blue solid and dotted lines represent our second-order (Eq. \ref{sec_amd}) and first-order (Eq. \ref{fir_amd}) AMD criteria, respectively. }
     \label{fig:q_dep}
    \end{figure}

\renewcommand{\thefigure}{B.\arabic{figure}}
\setcounter{figure}{1} 
    \begin{figure}
     \centering
     \includegraphics[width=0.5\textwidth]{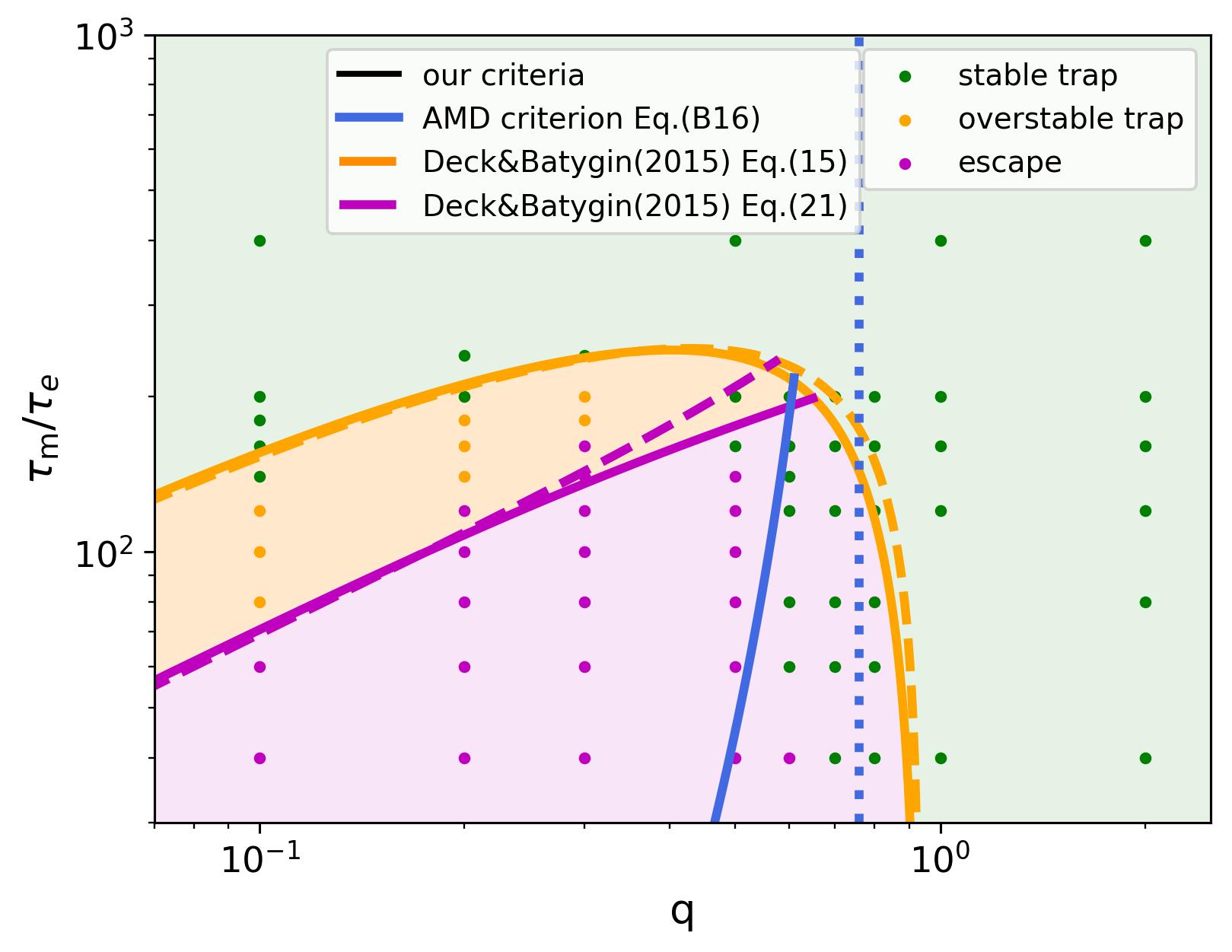}
     \caption{Same as Fig \ref{fig:q_dep}, but for the $3$:$2$ MMR.}
     \label{fig:q_dep_3_2}
    \end{figure}

\subsection*{B.4: Second-Order AMD criterion}
We expand the disturbing functions up to the second order in $e_{\rm i}$ and $e_{\rm o}$, which gives 
\begin{eqnarray}
    R_{\rm D}^{\rm sec} & = & K_{1} (e_{\rm i}^2 + e_{\rm o}^{2}) + K_{2}e_{\rm i}e_{\rm o}\cos{(\varpi_{\rm i}-\varpi_{\rm o})}, \\
    R_{\rm D}^{\rm res} & = & e_{\rm i}f_{\rm d,i}\cos{\varphi_{\rm i}} + e_{\rm o}f_{\rm d,o}\cos{\varphi_{\rm o}} \nonumber \\
    & & + [e_{\rm i}^{2}f_{3}\cos{\varphi_{3}} + e_{\rm i}e_{\rm o}f_{4}\cos{\varphi_{4}} + e_{\rm o}^{2}f_{5}\cos{\varphi_{5}}]\delta_{\rm j,2},  \\
    R_{\rm E} & = & -2e_{\rm o}\cos{\varphi_{\rm o}}\delta_{\rm j,2},  \\
    R_{\rm I} & = & -\frac{1}{2}e_{\rm o}\cos{\varphi_{\rm o}}\delta_{\rm j,2}, 
\end{eqnarray}
where $\varphi_{3} {=} 4\lambda_{\rm o} 
                  - 2\lambda_{\rm i}
                  -2\varpi_{\rm i}$,
$\varphi_{4} {=} 4\lambda_{\rm o} 
                  - 2\lambda_{\rm i}
                  -\varpi_{\rm i}
                  -\varpi_{\rm o}$,
$\varphi_{5} {=} 4\lambda_{\rm o} 
                  - 2\lambda_{\rm i}
                  -2\varpi_{\rm o}$.
Detailed expressions of $K_{1},K_{2},f_{3},f_{4}$ and $f_{5}$ can be found in Tables 8.1 and 8.4 of \cite{1999ssd..book.....M}. Lagrange’s equations can be written as
\begin{eqnarray}
    \dot{\varpi_{\rm i}} & = & \frac{\mu_{\rm o} n_{\rm i} \alpha}{e_{\rm i}} 
                               [ f_{\rm d,i}\cos{\varphi_{\rm i}}
                               + (2e_{\rm i}f_{3}\cos{\varphi_{3}} + e_{\rm o}f_{4}\cos{\varphi_{4}})\delta_{\rm j,2} \nonumber \\
                             &&  + 2K_{1}e_{\rm i} + K_{2}e_{\rm o}\cos{(\varpi_{\rm i}-\varpi_{\rm o})}] \label{sec_varpi_i}, \\ 
   \dot{\varpi_{\rm o}}  & = & \frac{\mu_{\rm i} n_{\rm o}}{e_{\rm o}}    
                               [ f_{\rm d,o}^{'}\cos{\varphi_{\rm o}}
                               + (2e_{\rm o}f_{5}\cos{\varphi_{5}} + e_{\rm i}f_{4}\cos{\varphi_{4}})\delta_{\rm j,2} \nonumber \\
                             &&  + 2K_{1}e_{\rm o} + K_{2}e_{\rm i}\cos{(\varpi_{\rm i}-\varpi_{\rm o})}]. \label{sec_varpi_o}
\end{eqnarray}

The angular momentum deficit (AMD) \citep{1997A&A...317L..75L} is given by
\begin{equation}
    AMD_{\rm i,o} \sim \mu_{i,o} \sqrt{a_{\rm i,o}} (1-\sqrt{1-e_{\rm i,o}^{2}}) \sim \mu_{\rm i,o} \sqrt{a_{\rm i,o}} e_{\rm i,o}^{2}.
\end{equation}
When the AMD of the inner planet exceeds that of the outer planet, the condition 
\begin{equation}
    e_{\rm o}^{2}/e_{\rm i}^{2} > q\sqrt{\alpha}, \label{AMD}
\end{equation}
is satisfied. Under this condition, the amplitude of oscillations following overstability is limited, and therefore the perturbations cannot grow sufficiently to cause escape. By substituting Eqs. \eqref{sec_varpi_i} and \eqref{sec_varpi_o} to eliminate either $e_{\rm i}$ and $e_{\rm o}$ under the condition $\dot{\varpi_{\rm i}}=\dot{\varpi_{\rm o}}$, we obtain the relationship between $e_{\rm i,o}$ and $q$. Assuming $e_{\rm i}$ takes the form:
    $e_{\rm i}^{2} = {\tau_e}/{j\tau_{\rm m}}$, 
and substituting this into the derived form, we obtain our second-order AMD criterion as
\begin{equation}
    \left(\frac{\tau_{\rm m}}{\tau_e}\right) < \frac{1}{j}\left[\frac{(f_{3}-K_{1}) + q^{'2}(f_{5}+K_{1})}{q^{'}f_{\rm d,o}^{'} + f_{\rm d,i}}\right]^{2}, \label{sec_amd}
\end{equation}
where $q^{'2} \equiv q\sqrt{\alpha}$. We can also obtain the first-order AMD criterion by substituting Eq. \eqref{e_i_eq} and \eqref{e_oeq} into Eq. \eqref{AMD}:
\begin{equation}
    q > \left(\frac{f_{\rm d,i}}{f_{\rm d,o}^{'}}\right)^2 \frac{1}{\sqrt{\alpha}}.
    \label{fir_amd}
\end{equation}
By comparison, the inclusion of the second-order terms induce a $\tau_{\rm m}/\tau_{e}$ dependency, which is absent in the first-order approximation. We plot our AMD criteria in Figs. \ref{fig:q_dep} and \ref{fig:q_dep_3_2} and find that the second-order AMD criterion exhibits improved agreement with the numerical results.

\subsection*{C: Extension to Unequal Eccentricity Damping Timescales}
We extend the framework to the planet pair systems with non-equal eccentricity damping. Equation \eqref{n_eq} can be written as 
\begin{equation}
\begin{aligned}
   &  -[ (j-1) \mu_{\rm o} n_{\rm i} \alpha
    + j \mu_{\rm i} n_{\rm o} ]
    ( e_{\rm i} f_{\rm d,i} \sin{\varphi_{\rm i}}
    + e_{\rm o} f_{\rm d,o}^{'} \sin{\varphi_{\rm o}} )\\
    &=  \frac{1}{\tau_{\rm m}}  + \frac{e_{\rm o}^2}{\tau_{e \rm,o}} 
    - \frac{e_{\rm i}^{2}}{\tau_{e \rm,i}}.
\end{aligned}
\end{equation}
The equilibrium eccentricities read
\begin{subequations}
\begin{eqnarray}
   e_{\rm i,eq}^{2}                      & = &\frac{\tau_{e,\rm i}}{\tau_{\rm m}}
                                           \frac{1}{1+q\sqrt{\alpha}}
                                           \frac{f_{\rm d,i}^{2}}{j f_{\rm d,i}^{2}+(j-1)\frac{\tau_{e \rm,i}}{\tau_{e \rm,o}}f_{\rm d,o}^{'2}q\sqrt{\alpha}}, \\
   e_{\rm o,eq}^{2}                      & = & \frac{\tau_{e,\rm i}}{\tau_{\rm m}}
                                           \frac{1}{1+q\sqrt{\alpha}}
                                           \frac{f_{\rm d,o}^{'2} q^{2}\alpha}{j f_{\rm d,i}^{2}+(j-1)\frac{\tau_{e \rm,i}}{\tau_{e \rm,o}}f_{\rm d,o}^{'2}q\sqrt{\alpha}}.
\end{eqnarray}
\end{subequations}
The key criteria can be re-formulated as
\begin{equation}
 \begin{cases}
  {\displaystyle   
 \tau_{e, \rm i}\tau_{\rm m} > \frac{1}{(1+q\sqrt{\alpha})}
                       \frac{1}{[j f_{\rm d,i}^{2} + (j-1)\frac{\tau_{e, \rm i}}{\tau_{e, \rm o}} f_{\rm d,o}^{',2} q\sqrt{\alpha}]}
                       \frac{1}{(\mu_{\rm o}n_{\rm i}\alpha)^{2}}   } \\
   {\displaystyle    }  
     \hfill  [\mbox{weak e-damping}],  \vspace{0.1cm}\\
 {\displaystyle       \tau_{\rm m} > \frac{1}{[ (j-1) \mu_{\rm o} n_{\rm i} \alpha
    + j \mu_{\rm i} n_{\rm o} ]} 
    \frac{[(j-1)^{2}\mu_{\rm o}n_{\rm i}^{2}\alpha + j^{2}\mu_{\rm i}n_{\rm o}^{2}]^{1/3}}{3^{1/3}(\mu_{\rm o}n_{\rm i}\alpha f_{\rm d,i}^{2} + \mu_{\rm i}n_{\rm o}f_{\rm d,o}^{'2})^{2/3}}  }\\
     {\displaystyle   } 
       \hfill  [\mbox{slow migration}],  \vspace{0.1cm}\\
 {\displaystyle   \frac{\tau_{\rm m}}{\tau_{e, \rm i}} > \left(\frac{3}{\mu_o}\right)^{2/3} h(q)\ \ \left[\frac{(j-1)^{2}}{f_{\rm d,i}\alpha} \right]^{2/3} (j-1)^{2/3} f(q)} \\
   {\displaystyle    }  
     \hfill  [\mbox{stability criterion}],  \vspace{0.1cm}\\
 {\displaystyle   \frac{\tau_{\rm m}}{\tau_{e, \rm i}} < \left(\frac{3}{\mu_o}\right)^{2/3} \frac{h(q)}{4}\ \ \left[ \frac{(j-1)^{2}+j^{2}q}{f_{\rm d,i}\alpha + f_{\rm d,o}^{'}q^{2}} \right]^{2/3} }\\
   {\displaystyle   } 
        \hfill  [\mbox{escape criterion}],
\end{cases}
\end{equation}
where  
\begin{eqnarray}
    h(q) & = & \frac{1}{(1+q\sqrt{\alpha})} 
               \frac{f_{\rm d,i}^{2}+f_{\rm d,o}^{'2}q^{2}\alpha}{j f_{\rm d,i}^{2}+(j-1)\frac{\tau_{e, \rm i}}{\tau_{e, \rm o}}f_{\rm d,o}^{'2}q\sqrt{\alpha}},\\
    f(q) & = & 1-\left(\frac{f_{\rm d,o}^{'}}{f_{\rm d,i}}\right)^{2}q^{2}\alpha.
\end{eqnarray}
These criteria have also been validated and compared with those of \cite{2015ApJ...810..119D} through simulations by incorporating different eccentricity dampings of two planets at $\tau_{e,\rm i}/\tau_{e,\rm o}{ =} 2$ (see Fig. \ref{fig:q_1_diff_te}). However, we also find that the analytical formulas increasingly deviate from the simulation results as the ratio $\tau_{e,i}/\tau_{e,o}$ increases, particularly for values greater than 10. We suspect that this is because, in our derivation, $\cos{\varphi_{\rm i}}{=}\cos{\varphi_{\rm o}}{=}1$ is assumed, whereas it can not hold for extreme eccentricity damping ratios. A detailed investigation is nevertheless beyond the scope of this work.

\renewcommand{\thefigure}{C.\arabic{figure}}
\setcounter{figure}{0} 
    \begin{figure}
     \centering
     \includegraphics[width=0.5\textwidth]{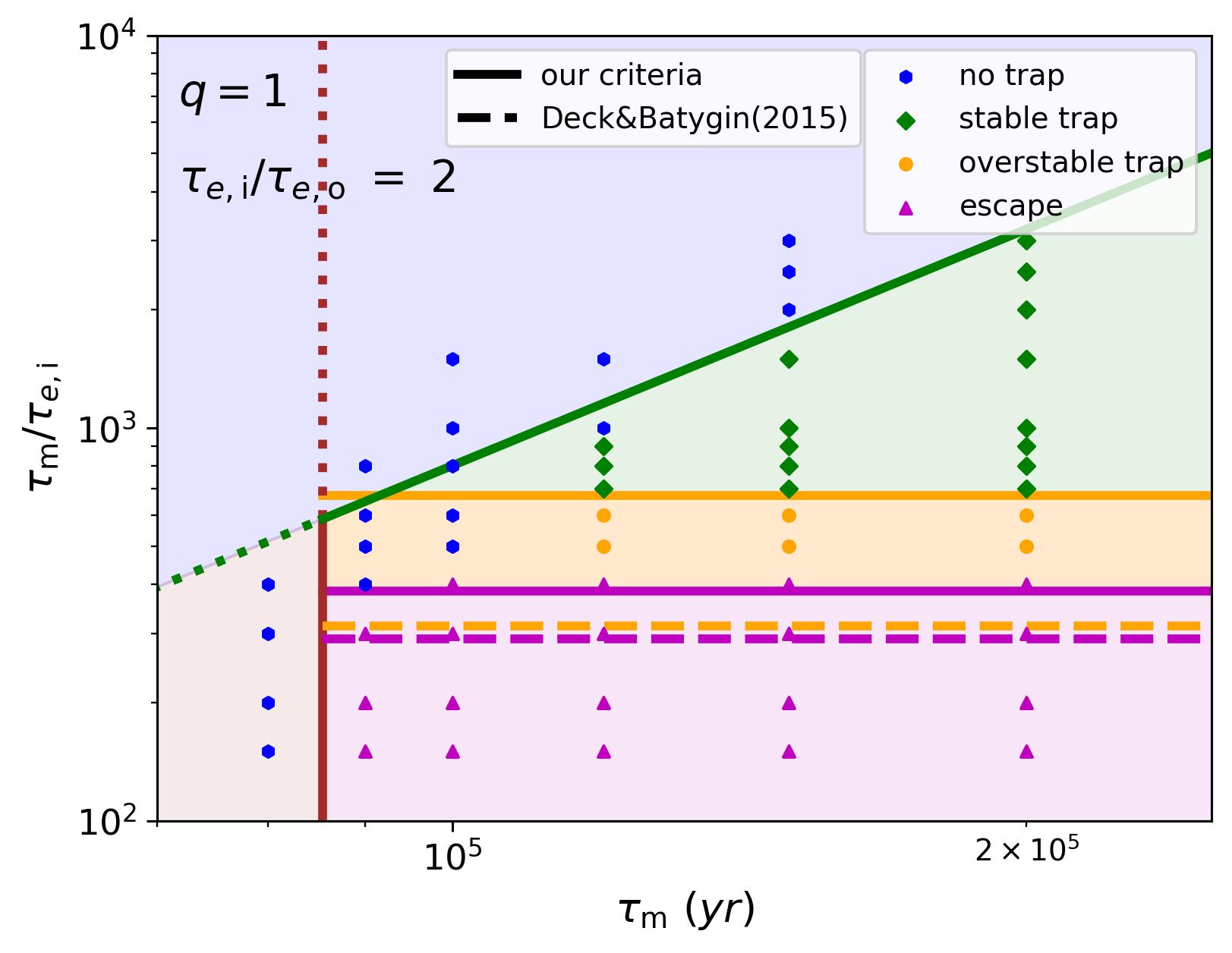}
     \caption{Same as Fig \ref{fig:q=1}, but $\tau_{e,\rm i}/\tau_{e,\rm o}=2$.}
     \label{fig:q_1_diff_te}
    \end{figure}

\end{document}